\documentclass[%
 reprint,
superscriptaddress,
 amsmath,amssymb,
 aps,
 prl,
]{revtex4-2}

\usepackage{graphicx}
\usepackage{dcolumn}
\usepackage{bm}
\usepackage{xcolor}
\usepackage[caption=false]{subfig}
\usepackage{multirow}

\begin{document}
\preprint{APS/123-QED}

\title{Search for an Excess of Electron Neutrino Interactions in MicroBooNE Using Multiple Final State Topologies}

\newcommand{\Bern}{Universit{\"a}t Bern, Bern CH-3012, Switzerland}
\newcommand{\BNL}{Brookhaven National Laboratory (BNL), Upton, NY, 11973, USA}
\newcommand{\UCSB}{University of California, Santa Barbara, CA, 93106, USA}
\newcommand{\Cambridge}{University of Cambridge, Cambridge CB3 0HE, United Kingdom}
\newcommand{\CIEMAT}{Centro de Investigaciones Energ\'{e}ticas, Medioambientales y Tecnol\'{o}gicas (CIEMAT), Madrid E-28040, Spain}
\newcommand{\Chicago}{University of Chicago, Chicago, IL, 60637, USA}
\newcommand{\Cincinnati}{University of Cincinnati, Cincinnati, OH, 45221, USA}
\newcommand{\CSU}{Colorado State University, Fort Collins, CO, 80523, USA}
\newcommand{\Columbia}{Columbia University, New York, NY, 10027, USA}
\newcommand{\Edinburgh}{University of Edinburgh, Edinburgh EH9 3FD, United Kingdom}
\newcommand{\FNAL}{Fermi National Accelerator Laboratory (FNAL), Batavia, IL 60510, USA}
\newcommand{\Granada}{Universidad de Granada, Granada E-18071, Spain}
\newcommand{\Harvard}{Harvard University, Cambridge, MA 02138, USA}
\newcommand{\IIT}{Illinois Institute of Technology (IIT), Chicago, IL 60616, USA}
\newcommand{\KSU}{Kansas State University (KSU), Manhattan, KS, 66506, USA}
\newcommand{\Lancaster}{Lancaster University, Lancaster LA1 4YW, United Kingdom}
\newcommand{\LANL}{Los Alamos National Laboratory (LANL), Los Alamos, NM, 87545, USA}
\newcommand{\Manchester}{The University of Manchester, Manchester M13 9PL, United Kingdom}
\newcommand{\MIT}{Massachusetts Institute of Technology (MIT), Cambridge, MA, 02139, USA}
\newcommand{\Michigan}{University of Michigan, Ann Arbor, MI, 48109, USA}
\newcommand{\Minnesota}{University of Minnesota, Minneapolis, MN, 55455, USA}
\newcommand{\NMSU}{New Mexico State University (NMSU), Las Cruces, NM, 88003, USA}
\newcommand{\Oxford}{University of Oxford, Oxford OX1 3RH, United Kingdom}
\newcommand{\Pitt}{University of Pittsburgh, Pittsburgh, PA, 15260, USA}
\newcommand{\Rutgers}{Rutgers University, Piscataway, NJ, 08854, USA}
\newcommand{\SLAC}{SLAC National Accelerator Laboratory, Menlo Park, CA, 94025, USA}
\newcommand{\SDSMT}{South Dakota School of Mines and Technology (SDSMT), Rapid City, SD, 57701, USA}
\newcommand{\Maine}{University of Southern Maine, Portland, ME, 04104, USA}
\newcommand{\Syracuse}{Syracuse University, Syracuse, NY, 13244, USA}
\newcommand{\TelAviv}{Tel Aviv University, Tel Aviv, Israel, 69978}
\newcommand{\Tennessee}{University of Tennessee, Knoxville, TN, 37996, USA}
\newcommand{\UTA}{University of Texas, Arlington, TX, 76019, USA}
\newcommand{\Tufts}{Tufts University, Medford, MA, 02155, USA}
\newcommand{\VTech}{Center for Neutrino Physics, Virginia Tech, Blacksburg, VA, 24061, USA}
\newcommand{\Warwick}{University of Warwick, Coventry CV4 7AL, United Kingdom}
\newcommand{\Yale}{Wright Laboratory, Department of Physics, Yale University, New Haven, CT, 06520, USA}

\affiliation{\Bern}
\affiliation{\BNL}
\affiliation{\UCSB}
\affiliation{\Cambridge}
\affiliation{\CIEMAT}
\affiliation{\Chicago}
\affiliation{\Cincinnati}
\affiliation{\CSU}
\affiliation{\Columbia}
\affiliation{\Edinburgh}
\affiliation{\FNAL}
\affiliation{\Granada}
\affiliation{\Harvard}
\affiliation{\IIT}
\affiliation{\KSU}
\affiliation{\Lancaster}
\affiliation{\LANL}
\affiliation{\Manchester}
\affiliation{\MIT}
\affiliation{\Michigan}
\affiliation{\Minnesota}
\affiliation{\NMSU}
\affiliation{\Oxford}
\affiliation{\Pitt}
\affiliation{\Rutgers}
\affiliation{\SLAC}
\affiliation{\SDSMT}
\affiliation{\Maine}
\affiliation{\Syracuse}
\affiliation{\TelAviv}
\affiliation{\Tennessee}
\affiliation{\UTA}
\affiliation{\Tufts}
\affiliation{\VTech}
\affiliation{\Warwick}
\affiliation{\Yale}

\author{P.~Abratenko} \affiliation{\Tufts} 
\author{R.~An} \affiliation{\IIT}
\author{J.~Anthony} \affiliation{\Cambridge}
\author{L.~Arellano} \affiliation{\Manchester}
\author{J.~Asaadi} \affiliation{\UTA}
\author{A.~Ashkenazi}\affiliation{\TelAviv}
\author{S.~Balasubramanian}\affiliation{\FNAL}
\author{B.~Baller} \affiliation{\FNAL}
\author{C.~Barnes} \affiliation{\Michigan}
\author{G.~Barr} \affiliation{\Oxford}
% ...

\collaboration{The MicroBooNE Collaboration}
\thanks{microboone\_info@fnal.gov}\noaffiliation

\begin{abstract}
We present a measurement of $\nu_e$ interactions from the Fermilab Booster Neutrino Beam using the MicroBooNE liquid argon time projection chamber to address the nature of the excess of low energy interactions observed by the MiniBooNE collaboration. Three independent $\nu_e$ searches are performed across multiple single electron final states, including an exclusive search for two-body scattering events with a single proton, a semi-inclusive search for pionless events, and a fully inclusive search for events containing all hadronic final states. With differing signal topologies, statistics, backgrounds, reconstruction algorithms, and analysis approaches, the results are found to be 
either consistent with or modestly lower than
the nominal $\nu_e$ rate expectations from the Booster Neutrino Beam and no excess of $\nu_e$ events is observed. 



\end{abstract}
\maketitle

MicroBooNE is the first liquid argon time projection chamber (LArTPC) to acquire high statistics samples of neutrino interactions on argon. Using this unique data set, MicroBooNE has pioneered a large body of results on neutrino interactions~\cite{wc_cc_inclusive,uB-NuMI-nue-nuebar,Abratenko_ccnp_2020,Abratenko_qe_2020,Abratenko_incl_2019,Adams_ccpi0_2019,Adams_mult_2019}, astrophysical~\cite{uB-atm-muon,uB-SN} and beyond the Standard Model physics~\cite{uB-higgs-portal,uB-NHL}, neutrino event reconstruction~\cite{2021calorimetric,WC-CR-rejection,WC-neutrino-selection,WC-3D-imaging,uB-SSNet,uB-MPID,uB-2track,uB-pi0-reco,uB-rejecting-cosmics,DL-pixel,uB-Pandora,DL-orig}, and detector properties~\cite{uB-diffusion,uB-SCE,uB-Efield,uB-calib,signal-processing-1,signal-processing-2,uB-CR-eff,uB-noise,uB-Michel,uB-MCS}. Here, we report the first measurement of electron neutrinos produced in the Fermilab Booster Neutrino Beamline (BNB) using the MicroBooNE detector. This multi-pronged search is aimed at investigating the as-yet unexplained low energy excess of electromagnetic activity observed by the MiniBooNE collaboration~\cite{Aguilar_Arevalo_2021}. 


Over the past decade, there has been a rich and evolving landscape of theoretical interpretations to explain the origin of the observed MiniBooNE excess, including standard processes~\cite{Giunti:2019sag} as well as new physics involving sterile neutrinos~\cite{Abazajian:2012ys,Bai:2015ztj}, dark sector portals~\cite{Bertuzzo:2018itn,Abdullahi:2020nyr,Alvarez-Ruso:2017hdm}, heavy neutral leptons~\cite{Ballett:2018ynz,Gninenko:2011xa}, non-standard Higgs physics~\cite{Dutta:2020scq,Asaadi:2017bhx,Abdallah:2020vgg,Abdallah:2020biq}, new particles produced in the beam~\cite{Chang:2021myh,Brdar:2020tle}, and mixed models of sterile neutrino oscillations and decay~\cite{Vergani:2021tgc,Fischer:2019fbw}. A number of scenarios have also been ruled out~\cite{Jordan:2018qiy}. Because of the variety of theoretical explanations and their possible signatures, MicroBooNE has developed three distinct $\nu_e$ searches targeting the MiniBooNE signal: an exclusive search for two-body $\nu_e$ charged current quasi-elastic (CCQE) scattering, a semi-inclusive search for pionless $\nu_e$ events, and an inclusive $\nu_e$ search containing any hadronic final state. Additionally, a companion single-photon-based search focused on radiative decays of the $\Delta$ resonance is reported elsewhere~\cite{gLEE_PRL}. This work capitalizes on the broad capabilities of a LArTPC to perform high purity measurements of electron neutrinos across multiple signal topologies and with significantly improved ability to distinguish whether an electromagnetic shower is electron or photon-induced compared to Cherenkov-based detectors such as MiniBooNE. 



The advantage of this particular probe of the MiniBooNE signal is that MicroBooNE is located in the same neutrino beamline and at roughly the same location as MiniBooNE, but uses an imaging detector capable of mm-scale spatial resolution and substantially lower energy detection thresholds for many particle types.  
The MicroBooNE LArTPC~\cite{Acciarri_2017,uB-CRT} itself contains 85 tons of liquid argon and is sited 72.5~m upstream of the MiniBooNE detector hall at a distance of 468.5~m from the BNB proton target.

The data used in this work are taken from an exposure of $7\times10^{20}$ protons on target (POT) collected in neutrino mode, 
a $93.7\%$ $\nu_\mu$ ($5.8\%$ $\overline{\nu}_\mu$) pure beam, 
from February 2016 to July 2018. These results represent an initial probe into electron neutrino production in the BNB using roughly half of the total data collected by MicroBooNE. Two different data streams are used in this analysis: an on-beam data sample triggered by BNB neutrino spills and an off-beam data sample taken during periods when no beam was received. The off-beam data sample is used for a direct data-based measurement of cosmic-induced backgrounds which are of importance given MicroBooNE's location near the surface. 

\begin{table*}[tbh!]
    \begin{tabular}{c|c|c} \hline
    $\nu_e$ Final State & Signal Constraints & Reconstruction Approach  \\ \hline
    $1e1p(0\pi)$ CCQE  & $\nu_\mu$ CCQE & Deep-Learning~\cite{DL_PRD} \\
    $1eN(\geq1)p0\pi$, $1e0p0\pi$ & $\nu_{\mu}$ CC & Pandora~\cite{PeLEE_PRD} \\
    $1eX$ & 
     $\nu_{\mu}$ CC, $\nu_{\mu}$ CC $\pi^0$, $\nu_\mu$ NC $\pi^0$ & Wire-Cell~\cite{WC_PRD} \\ \hline
   \end{tabular}
  \caption{Summary of signal definitions, signal-constraining data sets, and reconstruction approaches used for each of the three MicroBooNE $\nu_e$ searches. All samples require fully contained events with the exception of the $1eX$ analysis which additionally uses 
  both partially and fully contained $\nu_\mu$ CC samples as constraints.} 
  \label{tab:approach}
\end{table*}

While probing different event topologies with distinct event reconstruction methods, the three independent electron neutrino searches in MicroBooNE share several aspects in common. To simulate neutrino interactions in argon, the analyses rely on a \textsc{Geant4}-based~\cite{GEANT4:2002zbu} simulation of the neutrino beam~\cite{Aguilar_Arevalo_2009}, a variation of the \textsc{genie} v3 event generator~\cite{Andreopoulos_2010} specifically tuned to data that reflects our best knowledge of neutrino scattering in the BNB energy range~\cite{genie-tune-paper}, and a \textsc{Geant4}-based~\cite{GEANT4:2002zbu} detector simulation for particle propagation, with the processing of the charge response of the TPC and modeling of scintillation light implemented in the \textsc{LArSoft} framework~\cite{larsoft}. The data-driven detector simulation represents a significant upgrade from what has been historically available to model LArTPCs and incorporates pioneering work performed by MicroBooNE on wire signal processing~\cite{signal-processing-1,signal-processing-2}, noise removal~\cite{uB-noise}, electric field mapping~\cite{uB-Efield,uB-SCE}, and detector calibrations~\cite{uB-calib}. 

A common framework is additionally used to evaluate neutrino flux, neutrino cross section, and detector systematics. The evaluation of neutrino flux uncertainties is built on techniques developed by MiniBooNE~\cite{Aguilar_Arevalo_2009}. 
A total of more than 50 model parameters are varied within \textsc{genie} to assess uncertainties related to simulating neutrino interactions and final state effects in argon ~\cite{genie-tune-paper}. A full complement of LArTPC detector systematics are modeled, many with a novel data-driven technique built on comparisons of wire response in data and simulation~\cite{wire-mod-paper}. In addition, beam-related $\nu_e$ event predictions and uncertainties are further constrained 
through calculation of conditional means and variances
using data-driven measurements of neutrino charged current (CC) and neutral current (NC) interactions in MicroBooNE, with constraint samples tailored to each of the $\nu_e$ analysis approaches. Table~\ref{tab:approach} summarizes the signal definitions, constraint samples, and reconstruction approaches for the different analyses. 

Common approaches are also used for assessing agreement between observed and predicted $\nu_e$ samples and for hypothesis testing related to the presence or absence of an anomalous $\nu_e$ rate. In each search, the neutrino energy ($E_\nu$) is based on reconstructed final state particle energies~\cite{DL_PRD,PeLEE_PRD,WC_PRD}. Binned reconstructed $E_\nu$ distributions for data and simulation are then compared using a combined Neyman-Pearson test statistic, $\chi^2_{CNP}$~\cite{Ji:2019yca},
which approximates well to the Poisson-likelihood test statistic for low event numbers.
Similar statistical studies are performed to test a model of the $\nu_e$ event excess with an $E_\nu$ spectrum and normalization representative of that observed in the MiniBooNE experiment.  

A blind analysis scheme was adopted in which the signal region BNB $\nu_e$ data were only accessed after 
each analysis was completed.
Details on each $\nu_e$ analysis, including the optimization of purity and efficiency in each approach, data-based sidebands and simulated data sets used to validate the analyses, as well as data and Monte Carlo comparisons in a variety of kinematics can be found in Refs.~\cite{DL_PRD,PeLEE_PRD,WC_PRD}. The next three sections describe the strategy behind each of the $\nu_e$ searches and a summary of their results. 


\section{Two-Body $\nu_e$ CCQE Scattering ($1e1p$)}

An exclusive selection of $\nu_e$ candidates satisfying two-body CCQE kinematic constraints is performed in MicroBooNE with an analysis strategy focused around the use of deep-learning techniques~\cite{DL_PRD}.  
The CCQE process is predicted to dominate at low energies, with 77\% of $\nu_e$ events below 500~MeV (in true $E_\nu$) expected to interact via this channel in MicroBooNE.  
A strength of this selection is its clean final state topology, which simplifies event reconstruction and selection.  
The pure CCQE requirement also allows the use of kinematic variables for signal selection, such as proton momentum, transverse momentum, neutrino direction, momentum transfer, and Bjorken $x$.  
For this analysis, only fully-contained candidate $\nu_e$ and $\nu_{\mu}$ interactions with one reconstructed final state lepton and proton are selected.  
Final-state particle content and kinematics are reconstructed by applying a combination of conventional and deep-learning tools to prepared event images.  
Conventional tools are used to remove cosmic backgrounds~\cite{WC-3D-imaging}, identify candidate neutrino interaction  vertices~\cite{MicroBooNE:2020sar}, and reconstruct final-state particle candidates and their kinematics~\cite{MicroBooNE:2020sar, DL_shower}, while convolutional neural networks (CNNs) are used to differentiate track-like from shower-like image pixels~\cite{MicroBooNE:2020yze} and determine particle species contained in event images~\cite{MicroBooNE:2020hho}.  
After basic data quality selection criteria, Boosted Decision Tree (BDT) ensembles exploit 23 (16) kinematic and topological variables, including those described above, to collect a purified CCQE $\nu_e$ ($\nu_{\mu}$)  event set.  The CNN designed for particle identification is then used to select final candidates from this set.  

\begin{figure}[h]
\begin{center}
\includegraphics[width=1.0\columnwidth]{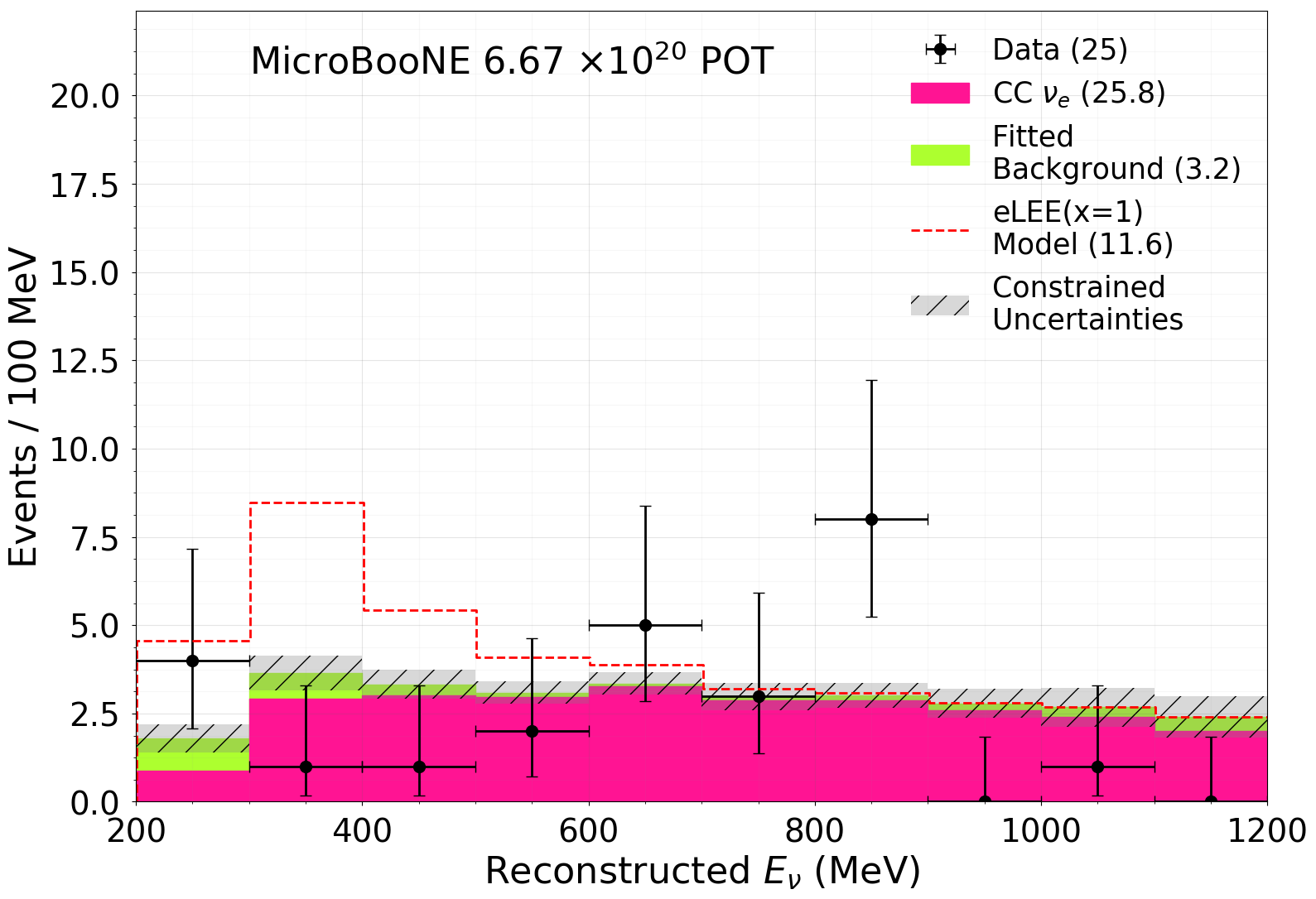}
\caption{Reconstructed neutrino energy for $1e1p$ CCQE candidate events in the deep-learning-based analysis. Backgrounds include contributions from cosmics and $\nu_{\mu}$ interactions. The $\nu_{e}$ prediction constrained using $\nu_{\mu}$ data is shown without (solid histogram) and with (red dotted) a model of the MiniBooNE low energy excess included (further detail in text).  Systematic uncertainties on the constrained prediction are shown as a hatched band.}
\label{fig:DL_ereco} 
\end{center}
\end{figure}

The results of the $\nu_{e}$ CCQE analysis are shown in Fig.~\ref{fig:DL_ereco}. This selection is predicted to produce a 75\% (75\%) pure sample of $\nu_e$ ($\nu_{\mu}$) CCQE events in the reconstructed $E_\nu$ range between 200 and 1200 MeV, with an efficiency of 6.6\% for all true $\nu_e$ CCQE interactions in the LArTPC active volume.  
The average $E_\nu$ energy resolution for selected $\nu_e$ events is estimated to be 16.5\%, with negligible bias predicted between true and reconstructed $E_\nu$. 
The $\nu_e$ CCQE prediction is constrained using a high-statistics $\nu_\mu$ CCQE candidate data set, which increases predicted $\nu_e$ counts by 6\% and reduces systematic uncertainties on the prediction by a factor of 2.0.
Statistical uncertainties dominate the final measurement.
The post-constraint prediction yields $29.0\,\pm\,5.2$ (stat.) $\pm\,1.9$\,(syst.) events in the full $E_\nu$ range given above (statistical errors follow the CNP formalism~\cite{Ji:2019yca}), with most events (89\%) predicted to arise from CCQE and non-CCQE $\nu_e$ interactions intrinsic to the beam. In the final selection, a total of 25 data candidates are observed in this range.  
The $\chi^2_\text{CNP}$ test statistic calculated between predicted and observed distributions is found to be 25.3 for the analysis in ten $E_{\nu}$ bins (where 6.9 units of $\chi^2_\text{CNP}$ result from a single bin at 850~MeV), leading to a $p$-value of 0.014.  
Below 500~MeV, the $\chi^2_\text{CNP}$ contribution is 7.9 for three $E_{\nu}$ bins, with data in two of the three bins falling slightly below predicted values.  

\section{Pionless $\nu_e$ Scattering ($1eNp0\pi$, $1e0p0\pi$)}\label{sec:pelee} 

A higher statistics search for pionless $\nu_e$ interactions that includes any number of protons in the final state uses the Pandora event reconstruction package~\cite{uB-Pandora}, which has been exercised over the years to produce a wide variety of MicroBooNE physics measurements~\cite{uB-NuMI-nue-nuebar,Abratenko_ccnp_2020,Abratenko_qe_2020,Abratenko_incl_2019,Adams_ccpi0_2019,Adams_mult_2019,uB-higgs-portal,uB-NHL}. The Pandora pattern recognition software, which reconstructs and classifies LArTPC events, is combined with specialized tools that further remove cosmic-ray background events as well as identify the different particles produced in a neutrino interaction \cite{ub_pid} and reconstruct their energies \cite{Adams_ccpi0_2019}. This search focuses on two exclusive channels with one electron and no pions in the final state: one with at least one visible proton ($1eNp0\pi$, $N\geq1$) and one with no visible protons ($1e0p0\pi$). A strength of this selection is that the two topologies combined exactly replicate the electron-like signal event signature in MiniBooNE.
This selection on fully contained events spanning neutrino energies from 10 to 2390 MeV provides an efficiency of 15$\%$ (9$\%$) with a purity of 80$\%$ (43$\%$) for $1eNp0\pi$ ($1e0p0\pi$) events. The typical energy resolution is $2\%$ for protons, $3\%$ for muons, and approximately $12\%$ for electrons, resulting in a predicted $E_\nu$ resolution of $15\%$ with $\sim5\%$ bias. To constrain neutrino flux and cross section uncertainties on the predicted intrinsic $\nu_e$ event rate, this analysis uses a high-statistics,
77$\%$ pure $\nu_{\mu}$ CC inclusive event sample~\cite{PeLEE_PRD}
and makes use of the cosmic-ray tagger detector system in MicroBooNE~\cite{uB-CRT} to further reduce cosmic backgrounds. This constraint reduces the systematic uncertainties in the $\nu_e$ selections by a factor of 1.7 and the result remains dominated by statistical uncertainties. 
This analysis is also validated using MicroBooNE data from the NuMI beam~\cite{Adamson:2015dkw} that provides a large number of $\nu_{e}$-argon interactions at a similar energy range as the BNB. 

\begin{figure*}[ht]
\begin{center}
\includegraphics[width=0.48\textwidth]{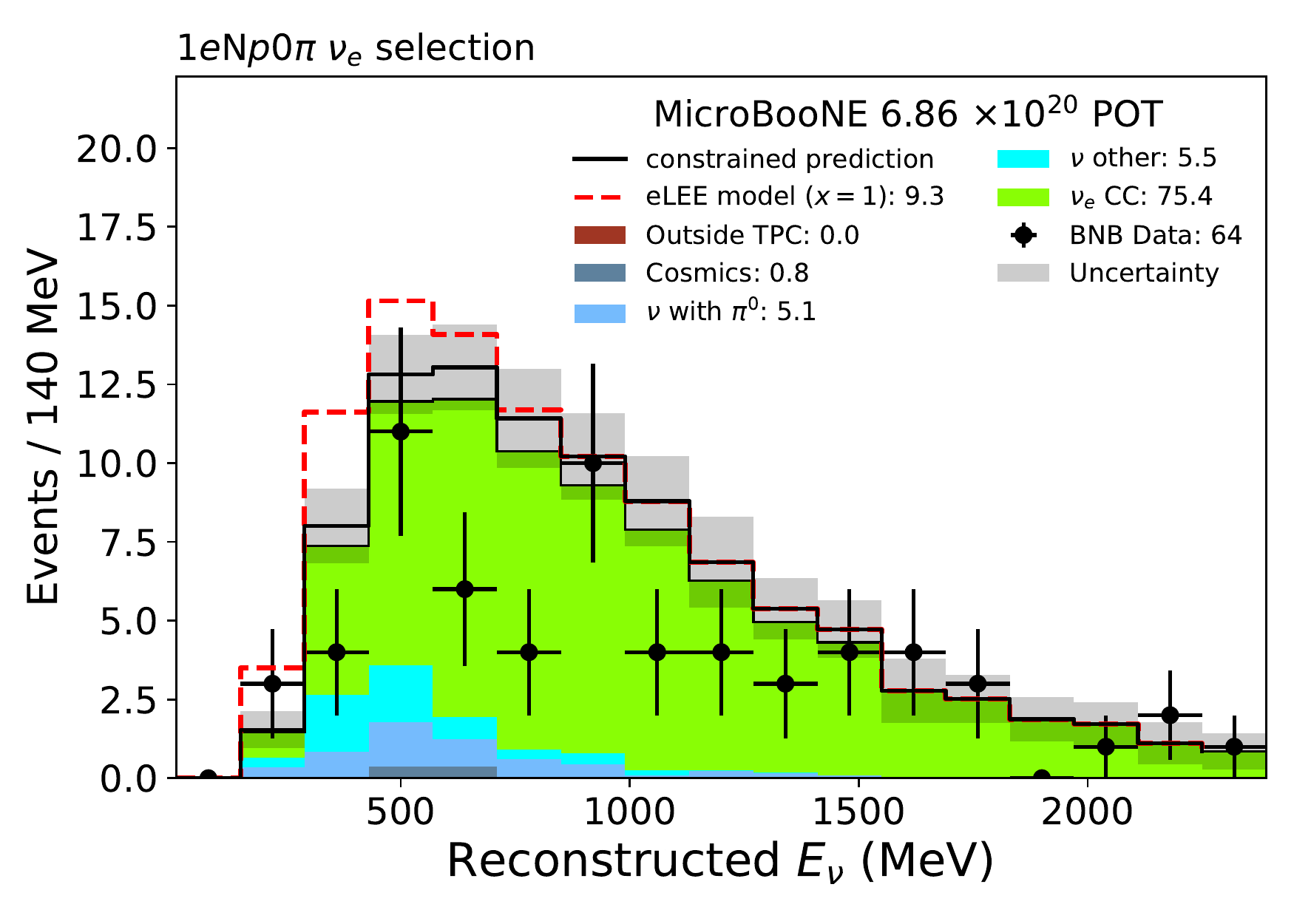}
\includegraphics[width=0.48\textwidth]{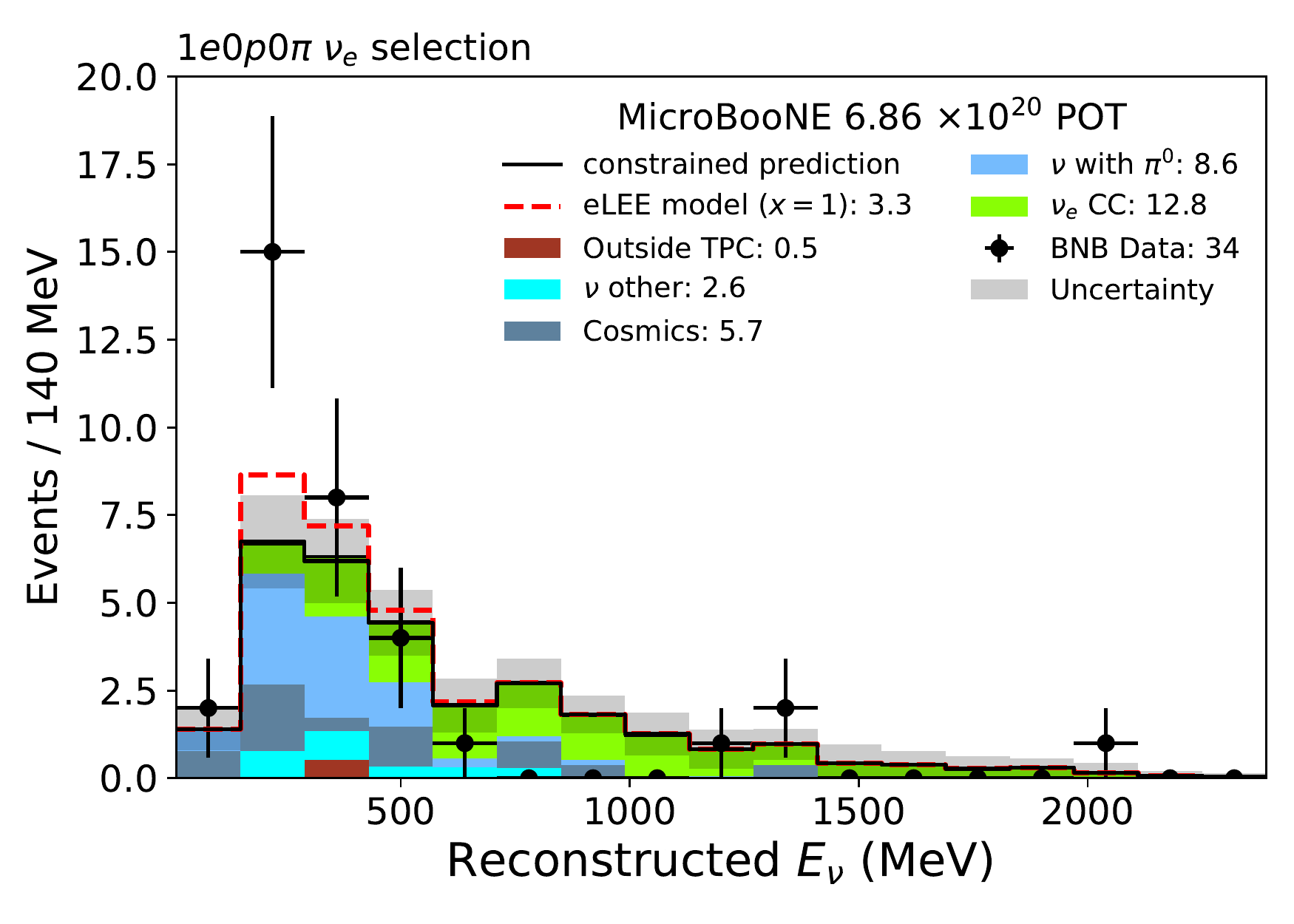}
\caption{Reconstructed neutrino energy for pionless $\nu_e$ candidate events in the Pandora-based analysis: $1eNp0\pi$ (left) and $1e0p0\pi$ (right). 
Each of the plots extend from 10 to 2390 MeV with a 140 MeV bin width.
The unconstrained number of predicted events is shown broken down by true interaction topology.  The constrained predictions using $\nu_\mu$ data are shown both with (red) and without (black) a model of the MiniBooNE low energy excess included (further detail in text). Systematic uncertainties on the constrained prediction are shown as a shaded band. 
}\label{fig:PeLEE_ereco} 
\end{center}
\end{figure*}

The results of the Pandora-based pionless $\nu_{e}$ analysis are shown in Fig.~\ref{fig:PeLEE_ereco}. For the $1eNp0\pi$ channel, $64$ $\nu_{e}$ data events are observed compared to $86.8\,\pm\,8.8$\,(stat)\,$\pm\,11.5$\,(syst) events expected (statistical errors follow the CNP formalism), in a reconstructed $E_\nu$ range between 10 and 2390 MeV. For the $1e0p0\pi$ channel, $34$ $\nu_{e}$ data events are observed compared to $30.2\,\pm\,5.6$\,(stat)\,$\pm\,4.3$\,(syst) events expected over that same energy range. The data are consistent with the prediction: in the region $150~\text{MeV} \leq E_\nu \leq 1550~\text{MeV}$ where the final statistical tests are performed, the $\chi^2_\text{CNP}/ndf$ (and associated $p$-values) relative to the nominal prediction are 14.9/10\,(0.194), 16.7.9/10\,(0.116), and 31.56/20\,(0.097) for the $1eNp0\pi$ channel, $1e0p0\pi$ channel, and both combined, respectively. As with the $1e1p$ CCQE search results, the data for the $1eNp0\pi$ channel fall slightly below prediction. For the $1e0p0\pi$ channel, the observed event count below 500 MeV is above prediction, albeit in a region with lower predicted $\nu_e$ purity.

\section{Inclusive $\nu_e$ Scattering ($1eX$)}\label{sec:WC} 

The highest statistics $\nu_e$ analysis in MicroBooNE searches inclusively for all possible hadronic final states such as the type of analyses that will be performed in the future wide-band Deep Underground Neutrino Experiment (DUNE) which will have larger contributions from additional inelastic scattering processes at higher energies. This analysis uses the Wire-Cell reconstruction paradigm~\cite{Qian:2018qbv} which forms three-dimensional images of particle-induced electron ionization tracks and showers via 1D wire position tomography. The 3D images are then processed by clustering algorithms and matched to light signals for cosmic rejection~\cite{MicroBooNE:2020vry,WC-neutrino-selection,WC-CR-rejection}, before a deep neural network~\cite{wc_pattern_recognition} is used to determine the neutrino candidate vertex. Finally, the events are characterized in terms of energy deposit, topology, and kinematics, for eventual event building, classification (e.g. $\nu_e$ CC, $\nu_\mu$ CC, $\pi^0$, cosmic), and neutrino energy reconstruction. The strengths of this approach are its high efficiency and high purity. After all selections, the predicted efficiency for selecting inclusive $\nu_e$ CC ($\nu_\mu$ CC) events is 46\% (68\%) with a purity of 82\% (92\%) for $0<E_\nu<2500$~MeV. For fully contained events, the predicted calorimetric-based $E_\nu$ resolution is 10--15\% (15--20\%) for $\nu_e$ CC ($\nu_\mu$ CC) events with $\sim$7\% (10\%) bias. In addition to the $\nu_\mu$ CC data samples, which include both fully and partially contained events in the detector, CC and NC interactions with a reconstructed $\pi^0$ serve as additional constraints for reducing systematic uncertainties and therefore maximizing sensitivity. A high statistics sample of $\nu_e$ events from the NuMI beam also serves to validate the analysis. The constraints reduce the fractional uncertainty on the predicted number of fully-contained $\nu_e$ CC events with reconstructed $E_\nu<600$~MeV by a factor of 3.5 relative to the unconstrained prediction. After constraints, the largest systematic uncertainties for fully-contained $\nu_e$ CC events 
are associated with limited Monte Carlo statistics associated with this rare event search, detector effects (mainly recombination and wire response), and neutrino cross section modeling~\cite{genie-tune-paper}. Compared to all systematic uncertainties, however, the statistical uncertainty on the data remains dominant. 

\begin{figure}[h]
\begin{center}
\includegraphics[width=1.0\columnwidth]{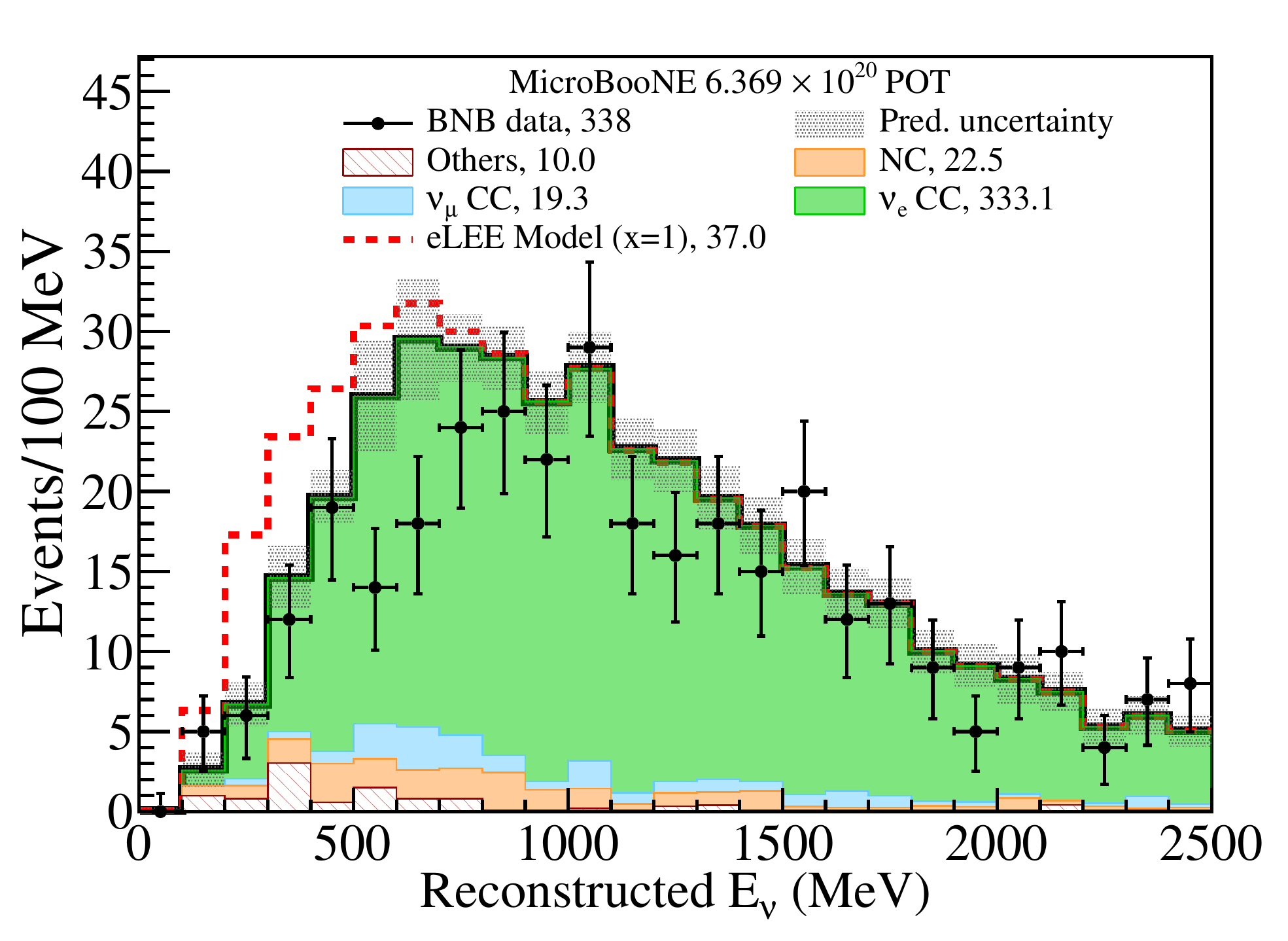}
\caption{\label{fig:WC_constrained} 
Reconstructed neutrino energy for inclusive $\nu_e$ candidate events in the Wire-Cell based analysis. The predicted event sample is dominated by $\nu_e$ events intrinsic to the beam (green) while the other background contributions are described in the legend. The constrained predictions are shown both with (red) and without (black) a model of the MiniBooNE low energy excess included (further detail in text). Systematic uncertainties on the constrained prediction are shown as a shaded band.}
\end{center}
\end{figure}

Fig.~\ref{fig:WC_constrained} shows the results of this inclusive $\nu_e$ search. The post-constraint $\nu_e$ CC inclusive analysis finds a modest deficit compared to the prediction: 56 (338) data events in $E_\nu<600$~MeV ($0<E_\nu<2500$~MeV) with $69.6\pm8.0~\mathrm{(stat.)}\pm5.0~\mathrm{(syst.)}$ ($384.9\pm19.2~\mathrm{(stat.)}\pm15.9~\mathrm{(syst.)}$) events expected. Good agreement is found between the data and the expectation from the BNB, with $\chi_{Pearson}^2/ndf=17.9/25$ and a corresponding $p$-value of $0.848$, across all energies.  Notably, agreement between the data and expectation is also apparent when the Wire-Cell inclusive event sample is studied in terms of its exclusive components, $1e0pX\pi$ and $1eNpX\pi$ ($X\geq0$), where these subsamples are further described in Ref.~\cite{WC_PRD}.

\begin{table*}[tbh!]
\parbox[t]{.4\textwidth}{
    \centering
    \hspace{0.35in}\textbf{Signal-enhanced region comparison} \\
    \vspace{0.5\baselineskip}
    \begin{tabular}{|r||c||c|c||c|}
    \hline
      & $1e1p$ & & & \\ 
      & CCQE & $1eNp0\pi$ & $1e0p0\pi$ & $1eX$ \\ \hline \hline
    $E_{\nu}$ (MeV) & {200-500} & {150-650} & {150-650} & {0-600} \rule{0pt}{2.6ex}\rule[-0.9ex]{0pt}{0pt}\\ 
    Predicted, & \multirow{2}{*}{$8.8\pm3.0$} & \multirow{2}{*}{$30.4\pm6.1$} & \multirow{2}{*}{$19.0\pm5.3$} & \multirow{2}{*}{$69.6\pm9.4$} \rule{0pt}{2.6ex}\\
    no eLEE & & & & \rule[-0.9ex]{0pt}{0pt}\\
    Predicted, & \multirow{2}{*}{$18.5\pm4.4$} & \multirow{2}{*}{$39.0\pm6.8$} & \multirow{2}{*}{$22.3\pm5.7$} & 
    \multirow{2}{*}{$104\pm12$} \rule{0pt}{2.6ex}\\
    w/ eLEE & & & & \rule[-0.9ex]{0pt}{0pt}\\
    Observed & 6 & 21 & 27 & 56 \rule{0pt}{2.6ex}\\ \hline 
   \end{tabular}
}
\hfill
\parbox[t]{.55\linewidth}{
    \textbf{Final fit results}\\
    \vspace{0.5\baselineskip}
    \begin{tabular}{|r||c||c|c||c|}
    \hline
     \textbf{} & $1e1p$ & & & \\ 
     \textbf{} & CCQE & $1eNp0\pi$ & $1e0p0\pi$ & $1eX$ \\ \hline \hline
    $E_{\nu}$ (MeV) & 200-1200 & 150-1550 & 150-1550 & 0-2500 \rule{0pt}{2.6ex}\rule[-0.9ex]{0pt}{0pt}\\ 
    $p~(\chi_{x=0}^2)$ & $1.4\times10^{-2}$ & 0.18 & 0.13 & 0.85 \rule{0pt}{2.6ex}\rule[-0.9ex]{0pt}{0pt}\\
    $p~(\Delta\chi^2 < \text{obs.})$,& \multirow{2}{*}{$1.6\times10^{-4}$} & \multirow{2}{*}{$2.1\times10^{-2}$} & \multirow{2}{*}{0.93} & \multirow{2}{*}{$9.0\times10^{-5}$} \rule{0pt}{2.6ex}\\
    w/ eLEE & & & & \rule[-0.9ex]{0pt}{0pt}\\
    \hline
    $x$ observed, $1\sigma$ & [0.00,0.08] & [0.00,0.41] & [1.91,8.10] & [0.00,0.22] \rule{0pt}{2.6ex}\rule[-0.9ex]{0pt}{0pt}\\
    $x$ observed, $2\sigma$ & [0.00,0.38] & [0.00,1.06] & [0.77,24.3] & [0.00,0.51] \rule{0pt}{2.6ex}\rule[-0.9ex]{0pt}{0pt}\\
    $x$ expected & \multirow{2}{*}{0.98} & \multirow{2}{*}{1.44} & \multirow{2}{*}{4.64} & \multirow{2}{*}{0.56} \rule{0pt}{2.6ex}\\
    upper limit, $2\sigma$ & & & & \\
        \hline
   \end{tabular}
  }
  \caption{\textit{Left}: Observed and predicted $\nu_e$ candidates in the signal-enhanced neutrino energy range predefined by each analysis prior to unblinding, in the absence ($x=0$) or presence ($x=1$) of a MiniBooNE-like $\nu_e$ event excess. This energy range is a subset of the full fit range, also chosen prior to unblinding. 
  Predicted events include the alternate-channel constraints of each analysis, and include statistical (following the CNP formalism) and constrained systematic uncertainties. \textit{Right}: Frequentist-derived $p$-values of the data observations compared to the prediction assuming no excess, $p~(\chi_{x=0}^2)$, and under a simple hypothesis test comparing an excess to no excess, $p (\Delta\chi^2 = \chi^2_{x=0} - \chi^2_{x=1} < \text{obs.})$, assuming the eLEE model ($x=1)$. Also quoted are the 1$\sigma$ and 2$\sigma$ confidence intervals for extracted signal strength $x$ over the full fit range and the expected 2$\sigma$ upper 
  endpoint of the interval
  on $x$ assuming no excess.} 
  \label{tab:rates}
\end{table*}




\section{Fit Results}\label{sec:fits} 
The three aforementioned analysis selections are not designed to be disjoint to each other, and there is an overlap in the selected events. Of the 25 events selected in the $1e1p$ CCQE analysis, 16 are selected in either the pionless or inclusive analysis. Of the 98 events selected across both pionless analysis selections, 46 are selected in the inclusive analysis.
All three analyses observe $\nu_e$ candidate event rates
in general agreement with or below the predicted rates.
Given the similar baseline and neutrino energies sampled by MicroBooNE and MiniBooNE, this picture appears to disfavor an interpretation of MiniBooNE's observed electron-like excess signature as arising purely from an anomalously high rate of charged current $\nu_e$ interactions.  
To more quantitatively address the comparison with the observed MiniBooNE data excess, all three analyses have performed statistical tests comparing datasets to a simple model of a MiniBooNE-like excess of $\nu_e$ interactions~\cite{MB-LEE-model}.

Using MiniBooNE simulation, a response matrix is constructed translating the 
true incident $E_\nu$ to reconstructed $E_\nu$
under a quasi-elastic assumption to the true incident $E_\nu$, accounting for detector response, acceptance, resolutions, event reconstruction, and selection efficiencies. 
Following 
a multi-dimensional
unfolding procedure~\cite{DAgostini:1994fjx} on the MiniBooNE observation~\cite{MiniBooNE:2018esg} and using only the statistical uncertainties on the MiniBooNE data and simulated events (to avoid any correlated uncertainties in flux and interaction models with MicroBooNE), MicroBooNE extracts an energy-dependent event rate of $\nu_e$ interactions. 
The resulting
scaling template, 
found to be robust against the number of unfolding iterations after an initial starting point corresponding to the MiniBooNE prediction,
is derived from the increase in event rate relative to the MiniBooNE prediction and then applied to simulated intrinsic $\nu_e$ events in MicroBooNE to form an ``eLEE" signal model, shown by the dashed lines in Figs.~\ref{fig:DL_ereco}--\ref{fig:WC_constrained}.  This scaling template varies only in true neutrino energy, thus the eLEE model otherwise assumes the same kinematics and final-state topologies as MicroBooNE's $\nu_e$ simulation -- additional kinematic information from the MiniBooNE excess result is not considered. The range of the eLEE signal model is  $200 <$ true $E_{\nu} < 800$~MeV, as the unfolding procedure does not consider data below 200 MeV in neutrino energy and finds no significant excess at higher energies.
 
 This simple model reproduces a median MiniBooNE electron-like excess to which MicroBooNE's results are compared, either by choosing a fixed excess normalization, $x$, matching MiniBooNE ($x$=1), or by treating $x$ as a free parameter to be extracted. 
 While the MiniBooNE uncertainties are not directly included in the eLEE model nor the statistical tests presented in this letter, 
 the reported significance of the excess from the MiniBooNE neutrino-mode data~\cite{Aguilar_Arevalo_2021}, $4.69\sigma$, translates to a $1\sigma$ confidence interval on the eLEE signal strength parameter of $1\pm0.21$, illustrating how the MiniBooNE excess would appear. More rigorous 
 comparisons of consistency with MiniBooNE in the future will need to consider correlated uncertainties in the neutrino flux and cross section models, as well as
 additional kinematic measurements. 
 

\begin{figure}[h]
\begin{center}
\includegraphics[width=1.0\columnwidth]{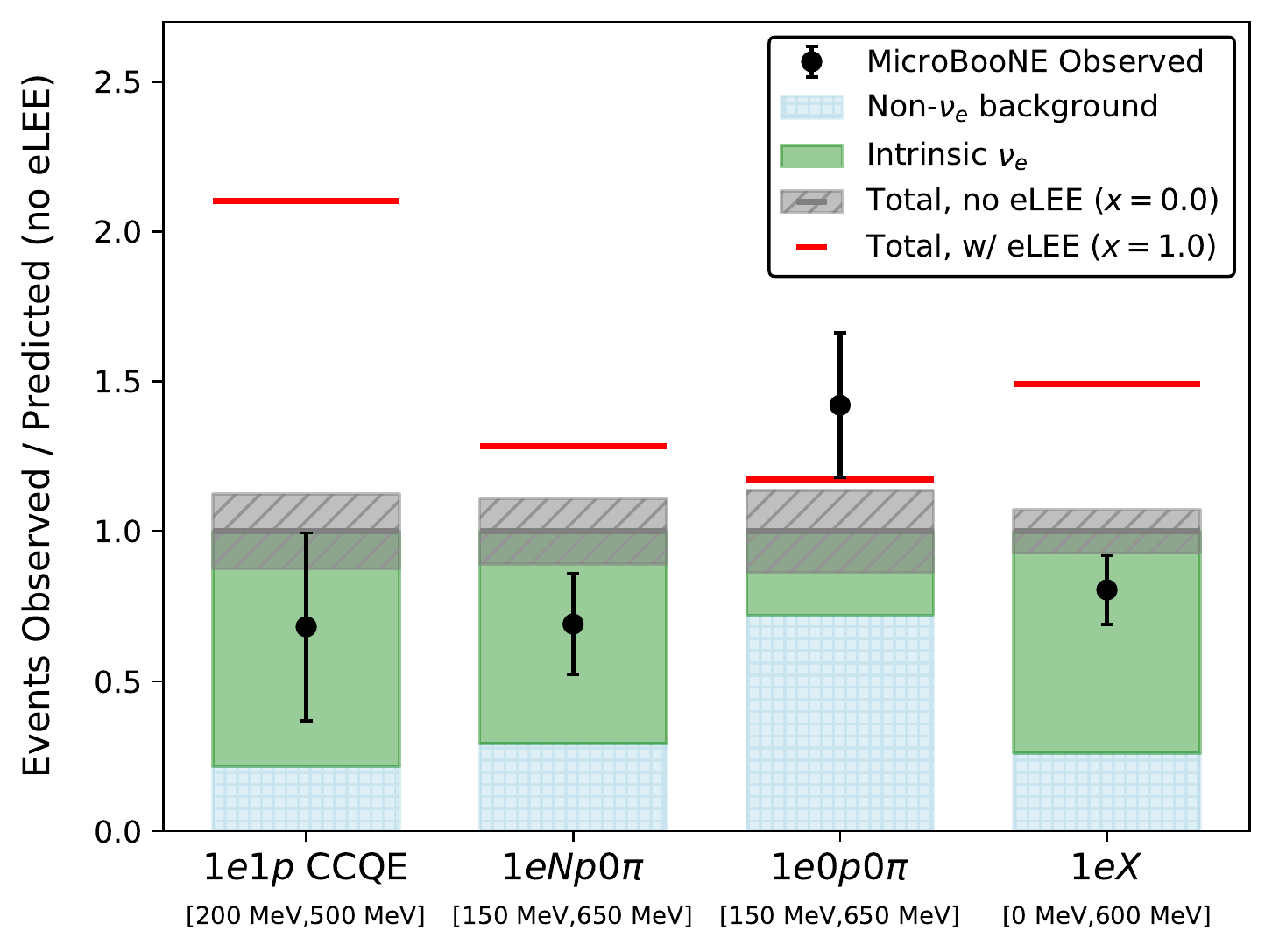}
\caption{Ratio of observed to predicted $\nu_e$ candidate events  -- assuming no eLEE -- in each analysis's signal-enhanced neutrino energy range (see Tab.~\ref{tab:rates}, left) 
with the relative contributions shown from non-$\nu_e$ backgrounds (light blue) and intrinsic $\nu_e$'s (green).
Statistical errors are shown on the observations (black), while systematic errors are shown around the prediction (gray). The expected ratio assuming the MiniBooNE-like eLEE signal model with its median signal strength is also shown (red).}
\label{fig:ratio_comparison} 
\end{center}
\end{figure}
 
Prior to unblinding of the data, each analysis defined a signal-enhanced low-energy region and determined the predicted number of events with and without the excess model in that region. Those predictions and the observed number of events are shown in Table~\ref{tab:rates} (left) and as ratios relative to the $x=0$ prediction in Fig.~\ref{fig:ratio_comparison}.

Each analysis performs two statistical analyses to test the signal hypothesis. First, a simple hypothesis test uses $\Delta\chi^2_\text{CNP} = \chi^2_{x=0}-\chi^2_{x=1}$ as a test statistic, comparing the observations to a frequentist $\Delta\chi^2_\text{CNP}$ distribution derived from model simulations assuming $x=0$ and $x=1$. The $p$-values corresponding to $\Delta\chi^2_\text{CNP}$ being less than the observed value assuming an eLEE (no eLEE) signal is $1.6\times10^{-4}$ ($0.02$), $0.021$ ($0.29$), $0.93$ ($0.98$), and $9.0\times10^{-5}$($0.33$) in the $1e1p$ CCQE, $1eNp0\pi$, $1e0p0\pi$, and $1eX$ selections, respectively. Each selection shows a strong preference for the absence of an electron-like MiniBooNE signal, with the exception of the $1e0p0\pi$ selection, driven by a data excess in the lowest energy bins, which also contain the highest contributions from non-$\nu_e$ backgrounds.

\begin{figure}[h]
\begin{center}
\includegraphics[width=1.0\columnwidth]{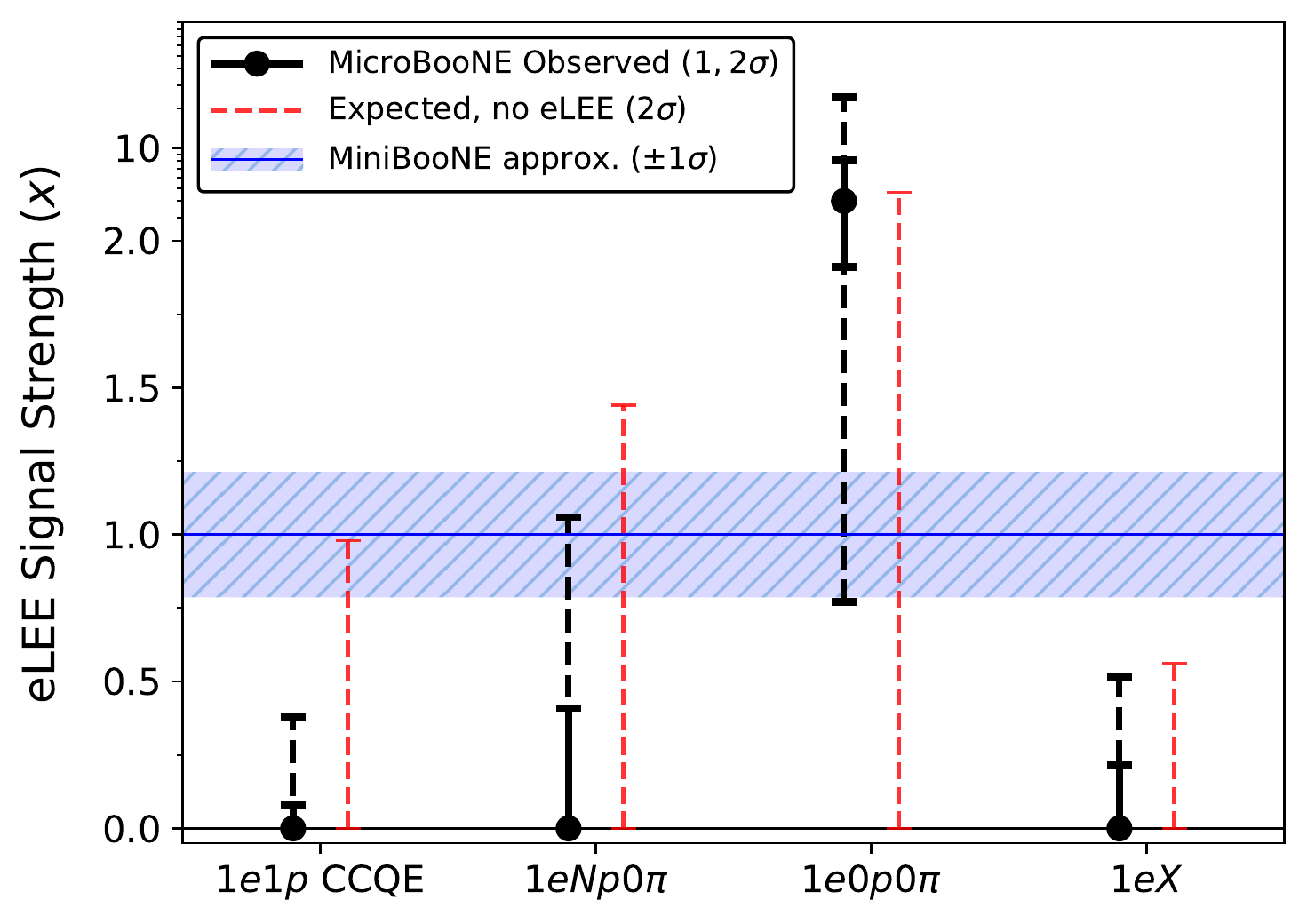}
\caption{Result of best-fit eLEE signal strength ($x$) in each analysis (black), along with the $1$ and $2\sigma$ confidence intervals (solid and dashed lines, respectively). The expected $2\sigma$ upper bound for each analysis, assuming no eLEE signal, is also shown (red). Signal strength values approximated from the MiniBooNE statistical and systematic errors (at $1\sigma$) are shown for comparison (blue). Note that the vertical scale is presented as linear from $x=0$ to $x=2$, while in logarithmic scale beyond that.}
\label{fig:x_fits} 
\end{center}
\end{figure}

Second, each analysis performs a nested hypothesis test where the eLEE signal strength $x$ is varied, with a lower bound constraint at $x=0$.
Each analysis independently finds a best-fit signal strength, $x_\text{min}$, by minimizing $\chi^2_\text{CNP}$. Following this, a test statistic defined as $\Delta\chi^2(x) = \chi^2_\text{CNP}(x)-\chi^2_\text{CNP}(x_\text{min})$ can be constructed for varying hypothetical signal strengths. A Feldman-Cousins method \cite{Feldman:1997qc} is used to construct confidence intervals around the best-fit signal strength, which are shown in Table~\ref{tab:rates} (right) and Fig.~\ref{fig:x_fits}. Consistent with the observed deficit of events at low reconstructed energies, the $1e1p$ CCQE,
$1eNp0\pi$, and $1eX$ selections each find a best fit signal strength of $x=0$, corresponding to the absence of an observed event excess, with $2\sigma$ upper bounds at $x<0.38$, $<1.06$, and $<0.51$, respectively. The expected $2\sigma$ upper bounds for these selections, assuming no signal, are shown in Table~\ref{tab:rates}. 
Consistent with the fact that in most analyses the observed number of events is less than the predicted number in the low energy regions, the measured upper endpoints of the $2\sigma$ interval are lower than expected.
The best-fit signal strength for the $1e0p0\pi$ selection is $x=4.0$, but with a wide confidence interval due to the low sensitivity of this channel. The best-fit signal strength for the $1eNp0\pi$ and $1e0p0\pi$ channels combined is $x=0.36$, with $x<1.86$ at the $2\sigma$ confidence level (and an expected upper bound where there is no signal at $x<1.37$), with more details in~\cite{PeLEE_PRD}.

\section{Conclusions}\label{sec:conclusions}
The MicroBooNE experiment has performed a set of inclusive and exclusive searches for $\nu_e$ CC events using $7 \times 10^{20}$ POT of Fermilab BNB neutrino-mode data, about half of the collected dataset, with each analysis considering a hypothesis for the nature of the MiniBooNE low-energy excess. This work and Ref.~\cite{gLEE_PRL} represent the first detailed study of this excess, noting that future MicroBooNE and SBN~\cite{MicroBooNE:2015bmn} measurements will continue to scrutinize the MiniBooNE results. The independent MicroBooNE search approaches have been led by distinct groups with each using a different fully automated event reconstruction software and common data-blindness scheme. All results reported here are unchanged since data unblinding. 

Afforded by the capabilities of the LArTPC technology to image various leptonic and hadronic final states, the searches all feature excellent signal identification and background rejection. In addition, the analyses use data-driven $\nu_e$ estimates constrained by high-statistics samples of $\pi^0$ and $\nu_\mu$ CC events. The expected event rate is dominated by intrinsic $\nu_e$ CC events originating from the beamline, rather than background events involving photons. Despite the near-surface location, cosmic rays represent a sub-dominant and usually negligible contribution to the backgrounds. 

No excess of low-energy $\nu_e$ candidates is observed, and the mutually compatible, statistics-limited measurements are either consistent with or modestly lower than the predictions for all $\nu_e$ event classes, including inclusive and exclusive hadronic final-states, and across all energies.
With the exception of the $1e0p0\pi$ selection which is the least sensitive to a simple model of the MiniBooNE low-energy excess, MicroBooNE rejects the hypothesis that $\nu_e$ CC interactions are fully responsible for that excess ($x=1$) at $>$97\% CL for both exclusive ($1e1p$ CCQE, $1eNp0\pi$) and inclusive ($1eX$) event classes.
Additionally, MicroBooNE disfavors generic $\nu_e$ interactions as the primary contributor to the excess, with a 1$\sigma$ (2$\sigma$) upper limit on the inclusive $\nu_e$ CC contribution to the excess of 22\% (51\%). While the MiniBooNE excess remains unexplained, our sensitive measurements are so far inconsistent with a $\nu_e$ interpretation of the excess.

\begin{acknowledgments}
\section{Acknowledgements}
This document was prepared by the MicroBooNE collaboration using the resources of the Fermi National Accelerator Laboratory (Fermilab), a U.S. Department of Energy, Office of Science, HEP User Facility. Fermilab is managed by Fermi Research Alliance, LLC (FRA), acting under Contract No. DE-AC02-07CH11359.  MicroBooNE is supported by the following: the U.S. Department of Energy, Office of Science, Offices of High Energy Physics and Nuclear Physics; the U.S. National Science Foundation; the Swiss National Science Foundation; the Science and Technology Facilities Council (STFC), part of the United Kingdom Research and Innovation; the Royal Society (United Kingdom); and The European Union’s Horizon 2020 Marie Sklodowska-Curie Actions. Additional support for the laser calibration system and cosmic-ray tagger was provided by the Albert Einstein Center for Fundamental Physics, Bern, Switzerland. We also acknowledge the contributions of technical and scientific staff to the design, construction, and operation of the MicroBooNE detector as well as the contributions of past collaborators to the development of MicroBooNE analyses, without whom this work would not have been possible.
\end{acknowledgments}
\bibliography{references}

\providecommand{\noopsort}[1]{}\providecommand{\singleletter}[1]{#1}%
\begin{thebibliography}{78}%
\makeatletter
\providecommand \@ifxundefined [1]{%
 \@ifx{#1\undefined}
}%
\providecommand \@ifnum [1]{%
 \ifnum #1\expandafter \@firstoftwo
 \else \expandafter \@secondoftwo
 \fi
}%
\providecommand \@ifx [1]{%
 \ifx #1\expandafter \@firstoftwo
 \else \expandafter \@secondoftwo
 \fi
}%
\providecommand \natexlab [1]{#1}%
\providecommand \enquote  [1]{``#1''}%
\providecommand \bibnamefont  [1]{#1}%
\providecommand \bibfnamefont [1]{#1}%
\providecommand \citenamefont [1]{#1}%
\providecommand \href@noop [0]{\@secondoftwo}%
\providecommand \href [0]{\begingroup \@sanitize@url \@href}%
\providecommand \@href[1]{\@@startlink{#1}\@@href}%
\providecommand \@@href[1]{\endgroup#1\@@endlink}%
\providecommand \@sanitize@url [0]{\catcode `\\12\catcode `\$12\catcode
  `\&12\catcode `\#12\catcode `\^12\catcode `\_12\catcode `\%12\relax}%
\providecommand \@@startlink[1]{}%
\providecommand \@@endlink[0]{}%
\providecommand \url  [0]{\begingroup\@sanitize@url \@url }%
\providecommand \@url [1]{\endgroup\@href {#1}{\urlprefix }}%
\providecommand \urlprefix  [0]{URL }%
\providecommand \Eprint [0]{\href }%
\providecommand \doibase [0]{http://dx.doi.org/}%
\providecommand \selectlanguage [0]{\@gobble}%
\providecommand \bibinfo  [0]{\@secondoftwo}%
\providecommand \bibfield  [0]{\@secondoftwo}%
\providecommand \translation [1]{[#1]}%
\providecommand \BibitemOpen [0]{}%
\providecommand \bibitemStop [0]{}%
\providecommand \bibitemNoStop [0]{.\EOS\space}%
\providecommand \EOS [0]{\spacefactor3000\relax}%
\providecommand \BibitemShut  [1]{\csname bibitem#1\endcsname}%
\let\auto@bib@innerbib\@empty
\bibitem [{\citenamefont {Abratenko}\ \emph
  {et~al.}({\natexlab{a}})\citenamefont {Abratenko} \emph
  {et~al.}}]{wc_cc_inclusive}%
  \BibitemOpen
  \bibfield  {author} {\bibinfo {author} {\bibfnamefont {P.}~\bibnamefont
  {Abratenko}} \emph {et~al.} (\bibinfo {collaboration} {MicroBooNE}),\
  }\Eprint {http://arxiv.org/abs/2110.14023} {arXiv:2110.14023 [hep-ex]}
  \BibitemShut {NoStop}%
\bibitem [{\citenamefont {Abratenko}\ \emph
  {et~al.}(2021{\natexlab{a}})\citenamefont {Abratenko} \emph
  {et~al.}}]{uB-NuMI-nue-nuebar}%
  \BibitemOpen
  \bibfield  {author} {\bibinfo {author} {\bibfnamefont {P.}~\bibnamefont
  {Abratenko}} \emph {et~al.} (\bibinfo {collaboration} {MicroBooNE}),\ }\href
  {\doibase 10.1103/PhysRevD.104.052002} {\bibfield  {journal} {\bibinfo
  {journal} {Phys. Rev. D}\ }\textbf {\bibinfo {volume} {104}},\ \bibinfo
  {pages} {052002} (\bibinfo {year} {2021}{\natexlab{a}})},\ \Eprint
  {http://arxiv.org/abs/2101.04228} {arXiv:2101.04228 [hep-ex]} \BibitemShut
  {NoStop}%
\bibitem [{\citenamefont {Abratenko}\ \emph
  {et~al.}(2020{\natexlab{a}})\citenamefont {Abratenko} \emph
  {et~al.}}]{Abratenko_ccnp_2020}%
  \BibitemOpen
  \bibfield  {author} {\bibinfo {author} {\bibfnamefont {P.}~\bibnamefont
  {Abratenko}} \emph {et~al.} (\bibinfo {collaboration} {MicroBooNE}),\ }\href
  {\doibase 10.1103/PhysRevD.102.112013} {\bibfield  {journal} {\bibinfo
  {journal} {Phys. Rev. D}\ }\textbf {\bibinfo {volume} {102}},\ \bibinfo
  {pages} {112013} (\bibinfo {year} {2020}{\natexlab{a}})},\ \Eprint
  {http://arxiv.org/abs/2010.02390} {arXiv:2010.02390 [hep-ex]} \BibitemShut
  {NoStop}%
\bibitem [{\citenamefont {Abratenko}\ \emph
  {et~al.}(2020{\natexlab{b}})\citenamefont {Abratenko} \emph
  {et~al.}}]{Abratenko_qe_2020}%
  \BibitemOpen
  \bibfield  {author} {\bibinfo {author} {\bibfnamefont {P.}~\bibnamefont
  {Abratenko}} \emph {et~al.} (\bibinfo {collaboration} {MicroBooNE}),\ }\href
  {\doibase 10.1103/PhysRevLett.125.201803} {\bibfield  {journal} {\bibinfo
  {journal} {Phys. Rev. Lett.}\ }\textbf {\bibinfo {volume} {125}},\ \bibinfo
  {pages} {201803} (\bibinfo {year} {2020}{\natexlab{b}})},\ \Eprint
  {http://arxiv.org/abs/2006.00108} {arXiv:2006.00108 [hep-ex]} \BibitemShut
  {NoStop}%
\bibitem [{\citenamefont {Abratenko}\ \emph {et~al.}(2019)\citenamefont
  {Abratenko} \emph {et~al.}}]{Abratenko_incl_2019}%
  \BibitemOpen
  \bibfield  {author} {\bibinfo {author} {\bibfnamefont {P.}~\bibnamefont
  {Abratenko}} \emph {et~al.} (\bibinfo {collaboration} {MicroBooNE}),\ }\href
  {\doibase 10.1103/PhysRevLett.123.131801} {\bibfield  {journal} {\bibinfo
  {journal} {Phys. Rev. Lett.}\ }\textbf {\bibinfo {volume} {123}},\ \bibinfo
  {pages} {131801} (\bibinfo {year} {2019})},\ \Eprint
  {http://arxiv.org/abs/1905.09694} {arXiv:1905.09694 [hep-ex]} \BibitemShut
  {NoStop}%
\bibitem [{\citenamefont {Adams}\ \emph
  {et~al.}(2019{\natexlab{a}})\citenamefont {Adams} \emph
  {et~al.}}]{Adams_ccpi0_2019}%
  \BibitemOpen
  \bibfield  {author} {\bibinfo {author} {\bibfnamefont {C.}~\bibnamefont
  {Adams}} \emph {et~al.} (\bibinfo {collaboration} {MicroBooNE}),\ }\href
  {\doibase 10.1103/PhysRevD.99.091102} {\bibfield  {journal} {\bibinfo
  {journal} {Phys. Rev. D}\ }\textbf {\bibinfo {volume} {99}},\ \bibinfo
  {pages} {091102} (\bibinfo {year} {2019}{\natexlab{a}})},\ \Eprint
  {http://arxiv.org/abs/1811.02700} {arXiv:1811.02700 [hep-ex]} \BibitemShut
  {NoStop}%
\bibitem [{\citenamefont {Adams}\ \emph
  {et~al.}(2019{\natexlab{b}})\citenamefont {Adams} \emph
  {et~al.}}]{Adams_mult_2019}%
  \BibitemOpen
  \bibfield  {author} {\bibinfo {author} {\bibfnamefont {C.}~\bibnamefont
  {Adams}} \emph {et~al.} (\bibinfo {collaboration} {MicroBooNE}),\ }\href
  {\doibase 10.1140/epjc/s10052-019-6742-3} {\bibfield  {journal} {\bibinfo
  {journal} {Eur. Phys. J. C}\ }\textbf {\bibinfo {volume} {79}},\ \bibinfo
  {pages} {248} (\bibinfo {year} {2019}{\natexlab{b}})},\ \Eprint
  {http://arxiv.org/abs/1805.06887} {arXiv:1805.06887 [hep-ex]} \BibitemShut
  {NoStop}%
\bibitem [{\citenamefont {Abratenko}\ \emph
  {et~al.}(2021{\natexlab{b}})\citenamefont {Abratenko} \emph
  {et~al.}}]{uB-atm-muon}%
  \BibitemOpen
  \bibfield  {author} {\bibinfo {author} {\bibfnamefont {P.}~\bibnamefont
  {Abratenko}} \emph {et~al.} (\bibinfo {collaboration} {MicroBooNE}),\ }\href
  {\doibase 10.1088/1748-0221/16/04/P04004} {\bibfield  {journal} {\bibinfo
  {journal} {JINST}\ }\textbf {\bibinfo {volume} {16}},\ \bibinfo {pages}
  {P04004} (\bibinfo {year} {2021}{\natexlab{b}})},\ \Eprint
  {http://arxiv.org/abs/2012.14324} {arXiv:2012.14324 [physics.ins-det]}
  \BibitemShut {NoStop}%
\bibitem [{\citenamefont {Abratenko}\ \emph
  {et~al.}(2021{\natexlab{c}})\citenamefont {Abratenko} \emph
  {et~al.}}]{uB-SN}%
  \BibitemOpen
  \bibfield  {author} {\bibinfo {author} {\bibfnamefont {P.}~\bibnamefont
  {Abratenko}} \emph {et~al.} (\bibinfo {collaboration} {MicroBooNE}),\ }\href
  {\doibase 10.1088/1748-0221/16/02/P02008} {\bibfield  {journal} {\bibinfo
  {journal} {JINST}\ }\textbf {\bibinfo {volume} {16}},\ \bibinfo {pages}
  {P02008} (\bibinfo {year} {2021}{\natexlab{c}})},\ \Eprint
  {http://arxiv.org/abs/2008.13761} {arXiv:2008.13761 [physics.ins-det]}
  \BibitemShut {NoStop}%
\bibitem [{\citenamefont {Abratenko}\ \emph
  {et~al.}(2021{\natexlab{d}})\citenamefont {Abratenko} \emph
  {et~al.}}]{uB-higgs-portal}%
  \BibitemOpen
  \bibfield  {author} {\bibinfo {author} {\bibfnamefont {P.}~\bibnamefont
  {Abratenko}} \emph {et~al.} (\bibinfo {collaboration} {MicroBooNE}),\ }\href
  {\doibase 10.1103/PhysRevLett.127.151803} {\bibfield  {journal} {\bibinfo
  {journal} {Phys. Rev. Lett.}\ }\textbf {\bibinfo {volume} {127}},\ \bibinfo
  {pages} {151803} (\bibinfo {year} {2021}{\natexlab{d}})},\ \Eprint
  {http://arxiv.org/abs/2106.00568} {arXiv:2106.00568 [hep-ex]} \BibitemShut
  {NoStop}%
\bibitem [{\citenamefont {Abratenko}\ \emph
  {et~al.}(2020{\natexlab{c}})\citenamefont {Abratenko} \emph
  {et~al.}}]{uB-NHL}%
  \BibitemOpen
  \bibfield  {author} {\bibinfo {author} {\bibfnamefont {P.}~\bibnamefont
  {Abratenko}} \emph {et~al.} (\bibinfo {collaboration} {MicroBooNE}),\ }\href
  {\doibase 10.1103/PhysRevD.101.052001} {\bibfield  {journal} {\bibinfo
  {journal} {Phys. Rev. D}\ }\textbf {\bibinfo {volume} {101}},\ \bibinfo
  {pages} {052001} (\bibinfo {year} {2020}{\natexlab{c}})},\ \Eprint
  {http://arxiv.org/abs/1911.10545} {arXiv:1911.10545 [hep-ex]} \BibitemShut
  {NoStop}%
\bibitem [{\citenamefont {Abratenko}\ \emph
  {et~al.}({\natexlab{b}})\citenamefont {Abratenko} \emph
  {et~al.}}]{2021calorimetric}%
  \BibitemOpen
  \bibfield  {author} {\bibinfo {author} {\bibfnamefont {P.}~\bibnamefont
  {Abratenko}} \emph {et~al.} (\bibinfo {collaboration} {MicroBooNE}),\
  }\Eprint {http://arxiv.org/abs/2109.02460} {arXiv:2109.02460
  [physics.ins-det]} \BibitemShut {NoStop}%
\bibitem [{\citenamefont {Abratenko}\ \emph
  {et~al.}(2021{\natexlab{e}})\citenamefont {Abratenko} \emph
  {et~al.}}]{WC-CR-rejection}%
  \BibitemOpen
  \bibfield  {author} {\bibinfo {author} {\bibfnamefont {P.}~\bibnamefont
  {Abratenko}} \emph {et~al.} (\bibinfo {collaboration} {MicroBooNE}),\ }\href
  {\doibase 10.1103/PhysRevApplied.15.064071} {\bibfield  {journal} {\bibinfo
  {journal} {Phys. Rev. Applied}\ }\textbf {\bibinfo {volume} {15}},\ \bibinfo
  {pages} {064071} (\bibinfo {year} {2021}{\natexlab{e}})},\ \Eprint
  {http://arxiv.org/abs/2101.05076} {arXiv:2101.05076 [physics.ins-det]}
  \BibitemShut {NoStop}%
\bibitem [{\citenamefont {Abratenko}\ \emph
  {et~al.}({\natexlab{c}})\citenamefont {Abratenko} \emph
  {et~al.}}]{WC-neutrino-selection}%
  \BibitemOpen
  \bibfield  {author} {\bibinfo {author} {\bibfnamefont {P.}~\bibnamefont
  {Abratenko}} \emph {et~al.} (\bibinfo {collaboration} {MicroBooNE}),\
  }\Eprint {http://arxiv.org/abs/2012.07928} {arXiv:2012.07928 [hep-ex]}
  \BibitemShut {NoStop}%
\bibitem [{\citenamefont {Abratenko}\ \emph
  {et~al.}(2021{\natexlab{f}})\citenamefont {Abratenko} \emph
  {et~al.}}]{WC-3D-imaging}%
  \BibitemOpen
  \bibfield  {author} {\bibinfo {author} {\bibfnamefont {P.}~\bibnamefont
  {Abratenko}} \emph {et~al.} (\bibinfo {collaboration} {MicroBooNE}),\ }\href
  {\doibase 10.1088/1748-0221/16/06/P06043} {\bibfield  {journal} {\bibinfo
  {journal} {JINST}\ }\textbf {\bibinfo {volume} {16}},\ \bibinfo {pages}
  {P06043} (\bibinfo {year} {2021}{\natexlab{f}})},\ \Eprint
  {http://arxiv.org/abs/2011.01375} {arXiv:2011.01375 [physics.ins-det]}
  \BibitemShut {NoStop}%
\bibitem [{\citenamefont {Abratenko}\ \emph
  {et~al.}(2021{\natexlab{g}})\citenamefont {Abratenko} \emph
  {et~al.}}]{uB-SSNet}%
  \BibitemOpen
  \bibfield  {author} {\bibinfo {author} {\bibfnamefont {P.}~\bibnamefont
  {Abratenko}} \emph {et~al.} (\bibinfo {collaboration} {MicroBooNE}),\ }\href
  {\doibase 10.1103/PhysRevD.103.052012} {\bibfield  {journal} {\bibinfo
  {journal} {Phys. Rev. D}\ }\textbf {\bibinfo {volume} {103}},\ \bibinfo
  {pages} {052012} (\bibinfo {year} {2021}{\natexlab{g}})},\ \Eprint
  {http://arxiv.org/abs/2012.08513} {arXiv:2012.08513 [physics.ins-det]}
  \BibitemShut {NoStop}%
\bibitem [{\citenamefont {Abratenko}\ \emph
  {et~al.}(2021{\natexlab{h}})\citenamefont {Abratenko} \emph
  {et~al.}}]{uB-MPID}%
  \BibitemOpen
  \bibfield  {author} {\bibinfo {author} {\bibfnamefont {P.}~\bibnamefont
  {Abratenko}} \emph {et~al.} (\bibinfo {collaboration} {MicroBooNE}),\ }\href
  {\doibase 10.1103/PhysRevD.103.092003} {\bibfield  {journal} {\bibinfo
  {journal} {Phys. Rev. D}\ }\textbf {\bibinfo {volume} {103}},\ \bibinfo
  {pages} {092003} (\bibinfo {year} {2021}{\natexlab{h}})},\ \Eprint
  {http://arxiv.org/abs/2010.08653} {arXiv:2010.08653 [hep-ex]} \BibitemShut
  {NoStop}%
\bibitem [{\citenamefont {Abratenko}\ \emph
  {et~al.}(2021{\natexlab{i}})\citenamefont {Abratenko} \emph
  {et~al.}}]{uB-2track}%
  \BibitemOpen
  \bibfield  {author} {\bibinfo {author} {\bibfnamefont {P.}~\bibnamefont
  {Abratenko}} \emph {et~al.} (\bibinfo {collaboration} {MicroBooNE}),\ }\href
  {\doibase 10.1088/1748-0221/16/02/P02017} {\bibfield  {journal} {\bibinfo
  {journal} {JINST}\ }\textbf {\bibinfo {volume} {16}},\ \bibinfo {pages}
  {P02017} (\bibinfo {year} {2021}{\natexlab{i}})},\ \Eprint
  {http://arxiv.org/abs/2002.09375} {arXiv:2002.09375 [physics.ins-det]}
  \BibitemShut {NoStop}%
\bibitem [{\citenamefont {Adams}\ \emph
  {et~al.}(2020{\natexlab{a}})\citenamefont {Adams} \emph
  {et~al.}}]{uB-pi0-reco}%
  \BibitemOpen
  \bibfield  {author} {\bibinfo {author} {\bibfnamefont {C.}~\bibnamefont
  {Adams}} \emph {et~al.} (\bibinfo {collaboration} {MicroBooNE}),\ }\href
  {\doibase 10.1088/1748-0221/15/02/P02007} {\bibfield  {journal} {\bibinfo
  {journal} {JINST}\ }\textbf {\bibinfo {volume} {15}},\ \bibinfo {pages}
  {P02007} (\bibinfo {year} {2020}{\natexlab{a}})},\ \Eprint
  {http://arxiv.org/abs/1910.02166} {arXiv:1910.02166 [hep-ex]} \BibitemShut
  {NoStop}%
\bibitem [{\citenamefont {Adams}\ \emph {et~al.}()\citenamefont {Adams} \emph
  {et~al.}}]{uB-rejecting-cosmics}%
  \BibitemOpen
  \bibfield  {author} {\bibinfo {author} {\bibfnamefont {C.}~\bibnamefont
  {Adams}} \emph {et~al.} (\bibinfo {collaboration} {MicroBooNE}),\ }\href
  {\doibase 10.1140/epjc/s10052-019-7184-7} {}\Eprint
  {http://arxiv.org/abs/1812.05679} {arXiv:1812.05679 [physics.ins-det]}
  \BibitemShut {NoStop}%
\bibitem [{\citenamefont {Adams}\ \emph
  {et~al.}(2019{\natexlab{c}})\citenamefont {Adams} \emph {et~al.}}]{DL-pixel}%
  \BibitemOpen
  \bibfield  {author} {\bibinfo {author} {\bibfnamefont {C.}~\bibnamefont
  {Adams}} \emph {et~al.} (\bibinfo {collaboration} {MicroBooNE}),\ }\href
  {\doibase 10.1103/PhysRevD.99.092001} {\bibfield  {journal} {\bibinfo
  {journal} {Phys. Rev. D}\ }\textbf {\bibinfo {volume} {99}},\ \bibinfo
  {pages} {092001} (\bibinfo {year} {2019}{\natexlab{c}})},\ \Eprint
  {http://arxiv.org/abs/1808.07269} {arXiv:1808.07269 [hep-ex]} \BibitemShut
  {NoStop}%
\bibitem [{\citenamefont {Acciarri}\ \emph {et~al.}(2018)\citenamefont
  {Acciarri} \emph {et~al.}}]{uB-Pandora}%
  \BibitemOpen
  \bibfield  {author} {\bibinfo {author} {\bibfnamefont {R.}~\bibnamefont
  {Acciarri}} \emph {et~al.} (\bibinfo {collaboration} {MicroBooNE}),\ }\href
  {\doibase 10.1140/epjc/s10052-017-5481-6} {\bibfield  {journal} {\bibinfo
  {journal} {Eur. Phys. J. C}\ }\textbf {\bibinfo {volume} {78}},\ \bibinfo
  {pages} {82} (\bibinfo {year} {2018})},\ \Eprint
  {http://arxiv.org/abs/1708.03135} {arXiv:1708.03135 [hep-ex]} \BibitemShut
  {NoStop}%
\bibitem [{\citenamefont {Acciarri}\ \emph
  {et~al.}(2017{\natexlab{a}})\citenamefont {Acciarri} \emph
  {et~al.}}]{DL-orig}%
  \BibitemOpen
  \bibfield  {author} {\bibinfo {author} {\bibfnamefont {R.}~\bibnamefont
  {Acciarri}} \emph {et~al.} (\bibinfo {collaboration} {MicroBooNE}),\ }\href
  {\doibase 10.1088/1748-0221/12/03/P03011} {\bibfield  {journal} {\bibinfo
  {journal} {JINST}\ }\textbf {\bibinfo {volume} {12}},\ \bibinfo {pages}
  {P03011} (\bibinfo {year} {2017}{\natexlab{a}})},\ \Eprint
  {http://arxiv.org/abs/1611.05531} {arXiv:1611.05531 [physics.ins-det]}
  \BibitemShut {NoStop}%
\bibitem [{\citenamefont {Abratenko}\ \emph
  {et~al.}({\natexlab{d}})\citenamefont {Abratenko} \emph
  {et~al.}}]{uB-diffusion}%
  \BibitemOpen
  \bibfield  {author} {\bibinfo {author} {\bibfnamefont {P.}~\bibnamefont
  {Abratenko}} \emph {et~al.} (\bibinfo {collaboration} {MicroBooNE}),\
  }\Eprint {http://arxiv.org/abs/2104.06551} {arXiv:2104.06551
  [physics.ins-det]} \BibitemShut {NoStop}%
\bibitem [{\citenamefont {Abratenko}\ \emph
  {et~al.}(2020{\natexlab{d}})\citenamefont {Abratenko} \emph
  {et~al.}}]{uB-SCE}%
  \BibitemOpen
  \bibfield  {author} {\bibinfo {author} {\bibfnamefont {P.}~\bibnamefont
  {Abratenko}} \emph {et~al.} (\bibinfo {collaboration} {MicroBooNE}),\ }\href
  {\doibase 10.1088/1748-0221/15/12/P12037} {\bibfield  {journal} {\bibinfo
  {journal} {JINST}\ }\textbf {\bibinfo {volume} {15}},\ \bibinfo {pages}
  {P12037} (\bibinfo {year} {2020}{\natexlab{d}})},\ \Eprint
  {http://arxiv.org/abs/2008.09765} {arXiv:2008.09765 [physics.ins-det]}
  \BibitemShut {NoStop}%
\bibitem [{\citenamefont {Adams}\ \emph
  {et~al.}(2020{\natexlab{b}})\citenamefont {Adams} \emph
  {et~al.}}]{uB-Efield}%
  \BibitemOpen
  \bibfield  {author} {\bibinfo {author} {\bibfnamefont {C.}~\bibnamefont
  {Adams}} \emph {et~al.} (\bibinfo {collaboration} {MicroBooNE}),\ }\href
  {\doibase 10.1088/1748-0221/15/07/P07010} {\bibfield  {journal} {\bibinfo
  {journal} {JINST}\ }\textbf {\bibinfo {volume} {15}},\ \bibinfo {pages}
  {P07010} (\bibinfo {year} {2020}{\natexlab{b}})},\ \Eprint
  {http://arxiv.org/abs/1910.01430} {arXiv:1910.01430 [physics.ins-det]}
  \BibitemShut {NoStop}%
\bibitem [{\citenamefont {Adams}\ \emph
  {et~al.}(2020{\natexlab{c}})\citenamefont {Adams} \emph {et~al.}}]{uB-calib}%
  \BibitemOpen
  \bibfield  {author} {\bibinfo {author} {\bibfnamefont {C.}~\bibnamefont
  {Adams}} \emph {et~al.} (\bibinfo {collaboration} {MicroBooNE}),\ }\href
  {\doibase 10.1088/1748-0221/15/03/P03022} {\bibfield  {journal} {\bibinfo
  {journal} {JINST}\ }\textbf {\bibinfo {volume} {15}},\ \bibinfo {pages}
  {P03022} (\bibinfo {year} {2020}{\natexlab{c}})},\ \Eprint
  {http://arxiv.org/abs/1907.11736} {arXiv:1907.11736 [physics.ins-det]}
  \BibitemShut {NoStop}%
\bibitem [{\citenamefont {Adams}\ \emph
  {et~al.}(2018{\natexlab{a}})\citenamefont {Adams} \emph
  {et~al.}}]{signal-processing-1}%
  \BibitemOpen
  \bibfield  {author} {\bibinfo {author} {\bibfnamefont {C.}~\bibnamefont
  {Adams}} \emph {et~al.} (\bibinfo {collaboration} {MicroBooNE}),\ }\href
  {\doibase 10.1088/1748-0221/13/07/P07006} {\bibfield  {journal} {\bibinfo
  {journal} {JINST}\ }\textbf {\bibinfo {volume} {13}},\ \bibinfo {pages}
  {P07006} (\bibinfo {year} {2018}{\natexlab{a}})},\ \Eprint
  {http://arxiv.org/abs/1802.08709} {arXiv:1802.08709 [physics.ins-det]}
  \BibitemShut {NoStop}%
\bibitem [{\citenamefont {Adams}\ \emph
  {et~al.}(2018{\natexlab{b}})\citenamefont {Adams} \emph
  {et~al.}}]{signal-processing-2}%
  \BibitemOpen
  \bibfield  {author} {\bibinfo {author} {\bibfnamefont {C.}~\bibnamefont
  {Adams}} \emph {et~al.} (\bibinfo {collaboration} {MicroBooNE}),\ }\href
  {\doibase 10.1088/1748-0221/13/07/P07007} {\bibfield  {journal} {\bibinfo
  {journal} {JINST}\ }\textbf {\bibinfo {volume} {13}},\ \bibinfo {pages}
  {P07007} (\bibinfo {year} {2018}{\natexlab{b}})},\ \Eprint
  {http://arxiv.org/abs/1804.02583} {arXiv:1804.02583 [physics.ins-det]}
  \BibitemShut {NoStop}%
\bibitem [{\citenamefont {Acciarri}\ \emph
  {et~al.}(2017{\natexlab{b}})\citenamefont {Acciarri} \emph
  {et~al.}}]{uB-CR-eff}%
  \BibitemOpen
  \bibfield  {author} {\bibinfo {author} {\bibfnamefont {R.}~\bibnamefont
  {Acciarri}} \emph {et~al.} (\bibinfo {collaboration} {MicroBooNE}),\ }\href
  {\doibase 10.1088/1748-0221/12/12/P12030} {\bibfield  {journal} {\bibinfo
  {journal} {JINST}\ }\textbf {\bibinfo {volume} {12}},\ \bibinfo {pages}
  {P12030} (\bibinfo {year} {2017}{\natexlab{b}})},\ \Eprint
  {http://arxiv.org/abs/1707.09903} {arXiv:1707.09903 [hep-ex]} \BibitemShut
  {NoStop}%
\bibitem [{\citenamefont {Acciarri}\ \emph
  {et~al.}(2017{\natexlab{c}})\citenamefont {Acciarri} \emph
  {et~al.}}]{uB-noise}%
  \BibitemOpen
  \bibfield  {author} {\bibinfo {author} {\bibfnamefont {R.}~\bibnamefont
  {Acciarri}} \emph {et~al.} (\bibinfo {collaboration} {MicroBooNE}),\ }\href
  {\doibase 10.1088/1748-0221/12/08/P08003} {\bibfield  {journal} {\bibinfo
  {journal} {JINST}\ }\textbf {\bibinfo {volume} {12}},\ \bibinfo {pages}
  {P08003} (\bibinfo {year} {2017}{\natexlab{c}})},\ \Eprint
  {http://arxiv.org/abs/1705.07341} {arXiv:1705.07341 [physics.ins-det]}
  \BibitemShut {NoStop}%
\bibitem [{\citenamefont {Acciarri}\ \emph
  {et~al.}(2017{\natexlab{d}})\citenamefont {Acciarri} \emph
  {et~al.}}]{uB-Michel}%
  \BibitemOpen
  \bibfield  {author} {\bibinfo {author} {\bibfnamefont {R.}~\bibnamefont
  {Acciarri}} \emph {et~al.} (\bibinfo {collaboration} {MicroBooNE}),\ }\href
  {\doibase 10.1088/1748-0221/12/09/P09014} {\bibfield  {journal} {\bibinfo
  {journal} {JINST}\ }\textbf {\bibinfo {volume} {12}},\ \bibinfo {pages}
  {P09014} (\bibinfo {year} {2017}{\natexlab{d}})},\ \Eprint
  {http://arxiv.org/abs/1704.02927} {arXiv:1704.02927 [physics.ins-det]}
  \BibitemShut {NoStop}%
\bibitem [{\citenamefont {Abratenko}\ \emph {et~al.}(2017)\citenamefont
  {Abratenko} \emph {et~al.}}]{uB-MCS}%
  \BibitemOpen
  \bibfield  {author} {\bibinfo {author} {\bibfnamefont {P.}~\bibnamefont
  {Abratenko}} \emph {et~al.} (\bibinfo {collaboration} {MicroBooNE}),\ }\href
  {\doibase 10.1088/1748-0221/12/10/P10010} {\bibfield  {journal} {\bibinfo
  {journal} {JINST}\ }\textbf {\bibinfo {volume} {12}},\ \bibinfo {pages}
  {P10010} (\bibinfo {year} {2017})},\ \Eprint
  {http://arxiv.org/abs/1703.06187} {arXiv:1703.06187 [physics.ins-det]}
  \BibitemShut {NoStop}%
\bibitem [{\citenamefont {Aguilar-Arevalo}\ \emph {et~al.}(2021)\citenamefont
  {Aguilar-Arevalo} \emph {et~al.}}]{Aguilar_Arevalo_2021}%
  \BibitemOpen
  \bibfield  {author} {\bibinfo {author} {\bibfnamefont {A.~A.}\ \bibnamefont
  {Aguilar-Arevalo}} \emph {et~al.} (\bibinfo {collaboration} {MiniBooNE}),\
  }\href {\doibase 10.1103/PhysRevD.103.052002} {\bibfield  {journal} {\bibinfo
   {journal} {Phys. Rev. D}\ }\textbf {\bibinfo {volume} {103}},\ \bibinfo
  {pages} {052002} (\bibinfo {year} {2021})},\ \Eprint
  {http://arxiv.org/abs/2006.16883} {arXiv:2006.16883 [hep-ex]} \BibitemShut
  {NoStop}%
\bibitem [{\citenamefont {Giunti}\ \emph {et~al.}(2020)\citenamefont {Giunti},
  \citenamefont {Ioannisian},\ and\ \citenamefont {Ranucci}}]{Giunti:2019sag}%
  \BibitemOpen
  \bibfield  {author} {\bibinfo {author} {\bibfnamefont {C.}~\bibnamefont
  {Giunti}}, \bibinfo {author} {\bibfnamefont {A.}~\bibnamefont {Ioannisian}},
  \ and\ \bibinfo {author} {\bibfnamefont {G.}~\bibnamefont {Ranucci}},\ }\href
  {\doibase 10.1007/JHEP11(2020)146} {\bibfield  {journal} {\bibinfo  {journal}
  {JHEP}\ }\textbf {\bibinfo {volume} {11}},\ \bibinfo {pages} {146} (\bibinfo
  {year} {2020})},\ \bibinfo {note} {[Erratum: JHEP 02, 078 (2021)]},\ \Eprint
  {http://arxiv.org/abs/1912.01524} {arXiv:1912.01524 [hep-ph]} \BibitemShut
  {NoStop}%
\bibitem [{\citenamefont {Abazajian}\ \emph {et~al.}()\citenamefont {Abazajian}
  \emph {et~al.}}]{Abazajian:2012ys}%
  \BibitemOpen
  \bibfield  {author} {\bibinfo {author} {\bibfnamefont {K.~N.}\ \bibnamefont
  {Abazajian}} \emph {et~al.},\ }\Eprint {http://arxiv.org/abs/1204.5379}
  {arXiv:1204.5379 [hep-ph]} \BibitemShut {NoStop}%
\bibitem [{\citenamefont {Bai}\ \emph {et~al.}(2016)\citenamefont {Bai},
  \citenamefont {Lu}, \citenamefont {Lu}, \citenamefont {Salvado},\ and\
  \citenamefont {Stefanek}}]{Bai:2015ztj}%
  \BibitemOpen
  \bibfield  {author} {\bibinfo {author} {\bibfnamefont {Y.}~\bibnamefont
  {Bai}}, \bibinfo {author} {\bibfnamefont {R.}~\bibnamefont {Lu}}, \bibinfo
  {author} {\bibfnamefont {S.}~\bibnamefont {Lu}}, \bibinfo {author}
  {\bibfnamefont {J.}~\bibnamefont {Salvado}}, \ and\ \bibinfo {author}
  {\bibfnamefont {B.~A.}\ \bibnamefont {Stefanek}},\ }\href {\doibase
  10.1103/PhysRevD.93.073004} {\bibfield  {journal} {\bibinfo  {journal} {Phys.
  Rev. D}\ }\textbf {\bibinfo {volume} {93}},\ \bibinfo {pages} {073004}
  (\bibinfo {year} {2016})},\ \Eprint {http://arxiv.org/abs/1512.05357}
  {arXiv:1512.05357 [hep-ph]} \BibitemShut {NoStop}%
\bibitem [{\citenamefont {Bertuzzo}\ \emph {et~al.}(2018)\citenamefont
  {Bertuzzo}, \citenamefont {Jana}, \citenamefont {Machado},\ and\
  \citenamefont {Zukanovich~Funchal}}]{Bertuzzo:2018itn}%
  \BibitemOpen
  \bibfield  {author} {\bibinfo {author} {\bibfnamefont {E.}~\bibnamefont
  {Bertuzzo}}, \bibinfo {author} {\bibfnamefont {S.}~\bibnamefont {Jana}},
  \bibinfo {author} {\bibfnamefont {P.~A.~N.}\ \bibnamefont {Machado}}, \ and\
  \bibinfo {author} {\bibfnamefont {R.}~\bibnamefont {Zukanovich~Funchal}},\
  }\href {\doibase 10.1103/PhysRevLett.121.241801} {\bibfield  {journal}
  {\bibinfo  {journal} {Phys. Rev. Lett.}\ }\textbf {\bibinfo {volume} {121}},\
  \bibinfo {pages} {241801} (\bibinfo {year} {2018})},\ \Eprint
  {http://arxiv.org/abs/1807.09877} {arXiv:1807.09877 [hep-ph]} \BibitemShut
  {NoStop}%
\bibitem [{\citenamefont {Abdullahi}\ \emph {et~al.}(2021)\citenamefont
  {Abdullahi}, \citenamefont {Hostert},\ and\ \citenamefont
  {Pascoli}}]{Abdullahi:2020nyr}%
  \BibitemOpen
  \bibfield  {author} {\bibinfo {author} {\bibfnamefont {A.}~\bibnamefont
  {Abdullahi}}, \bibinfo {author} {\bibfnamefont {M.}~\bibnamefont {Hostert}},
  \ and\ \bibinfo {author} {\bibfnamefont {S.}~\bibnamefont {Pascoli}},\ }\href
  {\doibase 10.1016/j.physletb.2021.136531} {\bibfield  {journal} {\bibinfo
  {journal} {Phys. Lett. B}\ }\textbf {\bibinfo {volume} {820}},\ \bibinfo
  {pages} {136531} (\bibinfo {year} {2021})},\ \Eprint
  {http://arxiv.org/abs/2007.11813} {arXiv:2007.11813 [hep-ph]} \BibitemShut
  {NoStop}%
\bibitem [{\citenamefont {Alvarez-Ruso}\ and\ \citenamefont
  {Saul-Sala}(2017)}]{Alvarez-Ruso:2017hdm}%
  \BibitemOpen
  \bibfield  {author} {\bibinfo {author} {\bibfnamefont {L.}~\bibnamefont
  {Alvarez-Ruso}}\ and\ \bibinfo {author} {\bibfnamefont {E.}~\bibnamefont
  {Saul-Sala}},\ }in\ \href@noop {} {\emph {\bibinfo {booktitle} {{Prospects in
  Neutrino Physics}}}}\ (\bibinfo {year} {2017})\ \Eprint
  {http://arxiv.org/abs/1705.00353} {arXiv:1705.00353 [hep-ph]} \BibitemShut
  {NoStop}%
\bibitem [{\citenamefont {Ballett}\ \emph {et~al.}(2019)\citenamefont
  {Ballett}, \citenamefont {Pascoli},\ and\ \citenamefont
  {Ross-Lonergan}}]{Ballett:2018ynz}%
  \BibitemOpen
  \bibfield  {author} {\bibinfo {author} {\bibfnamefont {P.}~\bibnamefont
  {Ballett}}, \bibinfo {author} {\bibfnamefont {S.}~\bibnamefont {Pascoli}}, \
  and\ \bibinfo {author} {\bibfnamefont {M.}~\bibnamefont {Ross-Lonergan}},\
  }\href {\doibase 10.1103/PhysRevD.99.071701} {\bibfield  {journal} {\bibinfo
  {journal} {Phys. Rev. D}\ }\textbf {\bibinfo {volume} {99}},\ \bibinfo
  {pages} {071701} (\bibinfo {year} {2019})},\ \Eprint
  {http://arxiv.org/abs/1808.02915} {arXiv:1808.02915 [hep-ph]} \BibitemShut
  {NoStop}%
\bibitem [{\citenamefont {Gninenko}(2011)}]{Gninenko:2011xa}%
  \BibitemOpen
  \bibfield  {author} {\bibinfo {author} {\bibfnamefont {S.~N.}\ \bibnamefont
  {Gninenko}},\ }\href {\doibase 10.1103/PhysRevD.83.093010} {\bibfield
  {journal} {\bibinfo  {journal} {Phys. Rev. D}\ }\textbf {\bibinfo {volume}
  {83}},\ \bibinfo {pages} {093010} (\bibinfo {year} {2011})},\ \Eprint
  {http://arxiv.org/abs/1101.4004} {arXiv:1101.4004 [hep-ex]} \BibitemShut
  {NoStop}%
\bibitem [{\citenamefont {Dutta}\ \emph {et~al.}(2020)\citenamefont {Dutta},
  \citenamefont {Ghosh},\ and\ \citenamefont {Li}}]{Dutta:2020scq}%
  \BibitemOpen
  \bibfield  {author} {\bibinfo {author} {\bibfnamefont {B.}~\bibnamefont
  {Dutta}}, \bibinfo {author} {\bibfnamefont {S.}~\bibnamefont {Ghosh}}, \ and\
  \bibinfo {author} {\bibfnamefont {T.}~\bibnamefont {Li}},\ }\href {\doibase
  10.1103/PhysRevD.102.055017} {\bibfield  {journal} {\bibinfo  {journal}
  {Phys. Rev. D}\ }\textbf {\bibinfo {volume} {102}},\ \bibinfo {pages}
  {055017} (\bibinfo {year} {2020})},\ \Eprint
  {http://arxiv.org/abs/2006.01319} {arXiv:2006.01319 [hep-ph]} \BibitemShut
  {NoStop}%
\bibitem [{\citenamefont {Asaadi}\ \emph {et~al.}(2018)\citenamefont {Asaadi},
  \citenamefont {Church}, \citenamefont {Guenette}, \citenamefont {Jones},\
  and\ \citenamefont {Szelc}}]{Asaadi:2017bhx}%
  \BibitemOpen
  \bibfield  {author} {\bibinfo {author} {\bibfnamefont {J.}~\bibnamefont
  {Asaadi}}, \bibinfo {author} {\bibfnamefont {E.}~\bibnamefont {Church}},
  \bibinfo {author} {\bibfnamefont {R.}~\bibnamefont {Guenette}}, \bibinfo
  {author} {\bibfnamefont {B.~J.~P.}\ \bibnamefont {Jones}}, \ and\ \bibinfo
  {author} {\bibfnamefont {A.~M.}\ \bibnamefont {Szelc}},\ }\href {\doibase
  10.1103/PhysRevD.97.075021} {\bibfield  {journal} {\bibinfo  {journal} {Phys.
  Rev. D}\ }\textbf {\bibinfo {volume} {97}},\ \bibinfo {pages} {075021}
  (\bibinfo {year} {2018})},\ \Eprint {http://arxiv.org/abs/1712.08019}
  {arXiv:1712.08019 [hep-ph]} \BibitemShut {NoStop}%
\bibitem [{\citenamefont {Abdallah}\ \emph {et~al.}(2021)\citenamefont
  {Abdallah}, \citenamefont {Gandhi},\ and\ \citenamefont
  {Roy}}]{Abdallah:2020vgg}%
  \BibitemOpen
  \bibfield  {author} {\bibinfo {author} {\bibfnamefont {W.}~\bibnamefont
  {Abdallah}}, \bibinfo {author} {\bibfnamefont {R.}~\bibnamefont {Gandhi}}, \
  and\ \bibinfo {author} {\bibfnamefont {S.}~\bibnamefont {Roy}},\ }\href
  {\doibase 10.1103/PhysRevD.104.055028} {\bibfield  {journal} {\bibinfo
  {journal} {Phys. Rev. D}\ }\textbf {\bibinfo {volume} {104}},\ \bibinfo
  {pages} {055028} (\bibinfo {year} {2021})},\ \Eprint
  {http://arxiv.org/abs/2010.06159} {arXiv:2010.06159 [hep-ph]} \BibitemShut
  {NoStop}%
\bibitem [{\citenamefont {Abdallah}\ \emph {et~al.}(2020)\citenamefont
  {Abdallah}, \citenamefont {Gandhi},\ and\ \citenamefont
  {Roy}}]{Abdallah:2020biq}%
  \BibitemOpen
  \bibfield  {author} {\bibinfo {author} {\bibfnamefont {W.}~\bibnamefont
  {Abdallah}}, \bibinfo {author} {\bibfnamefont {R.}~\bibnamefont {Gandhi}}, \
  and\ \bibinfo {author} {\bibfnamefont {S.}~\bibnamefont {Roy}},\ }\href
  {\doibase 10.1007/JHEP12(2020)188} {\bibfield  {journal} {\bibinfo  {journal}
  {JHEP}\ }\textbf {\bibinfo {volume} {12}},\ \bibinfo {pages} {188} (\bibinfo
  {year} {2020})},\ \Eprint {http://arxiv.org/abs/2006.01948} {arXiv:2006.01948
  [hep-ph]} \BibitemShut {NoStop}%
\bibitem [{\citenamefont {Chang}\ \emph {et~al.}(2021)\citenamefont {Chang},
  \citenamefont {Chen}, \citenamefont {Ho},\ and\ \citenamefont
  {Tseng}}]{Chang:2021myh}%
  \BibitemOpen
  \bibfield  {author} {\bibinfo {author} {\bibfnamefont {C.-H.~V.}\
  \bibnamefont {Chang}}, \bibinfo {author} {\bibfnamefont {C.-R.}\ \bibnamefont
  {Chen}}, \bibinfo {author} {\bibfnamefont {S.-Y.}\ \bibnamefont {Ho}}, \ and\
  \bibinfo {author} {\bibfnamefont {S.-Y.}\ \bibnamefont {Tseng}},\ }\href
  {\doibase 10.1103/PhysRevD.104.015030} {\bibfield  {journal} {\bibinfo
  {journal} {Phys. Rev. D}\ }\textbf {\bibinfo {volume} {104}},\ \bibinfo
  {pages} {015030} (\bibinfo {year} {2021})},\ \Eprint
  {http://arxiv.org/abs/2102.05012} {arXiv:2102.05012 [hep-ph]} \BibitemShut
  {NoStop}%
\bibitem [{\citenamefont {Brdar}\ \emph {et~al.}(2021)\citenamefont {Brdar},
  \citenamefont {Fischer},\ and\ \citenamefont {Smirnov}}]{Brdar:2020tle}%
  \BibitemOpen
  \bibfield  {author} {\bibinfo {author} {\bibfnamefont {V.}~\bibnamefont
  {Brdar}}, \bibinfo {author} {\bibfnamefont {O.}~\bibnamefont {Fischer}}, \
  and\ \bibinfo {author} {\bibfnamefont {A.~Y.}\ \bibnamefont {Smirnov}},\
  }\href {\doibase 10.1103/PhysRevD.103.075008} {\bibfield  {journal} {\bibinfo
   {journal} {Phys. Rev. D}\ }\textbf {\bibinfo {volume} {103}},\ \bibinfo
  {pages} {075008} (\bibinfo {year} {2021})},\ \Eprint
  {http://arxiv.org/abs/2007.14411} {arXiv:2007.14411 [hep-ph]} \BibitemShut
  {NoStop}%
\bibitem [{\citenamefont {Vergani}\ \emph {et~al.}()\citenamefont {Vergani},
  \citenamefont {Kamp}, \citenamefont {Diaz}, \citenamefont {Arg\"uelles},
  \citenamefont {Conrad}, \citenamefont {Shaevitz},\ and\ \citenamefont
  {Uchida}}]{Vergani:2021tgc}%
  \BibitemOpen
  \bibfield  {author} {\bibinfo {author} {\bibfnamefont {S.}~\bibnamefont
  {Vergani}}, \bibinfo {author} {\bibfnamefont {N.~W.}\ \bibnamefont {Kamp}},
  \bibinfo {author} {\bibfnamefont {A.}~\bibnamefont {Diaz}}, \bibinfo {author}
  {\bibfnamefont {C.~A.}\ \bibnamefont {Arg\"uelles}}, \bibinfo {author}
  {\bibfnamefont {J.~M.}\ \bibnamefont {Conrad}}, \bibinfo {author}
  {\bibfnamefont {M.~H.}\ \bibnamefont {Shaevitz}}, \ and\ \bibinfo {author}
  {\bibfnamefont {M.~A.}\ \bibnamefont {Uchida}},\ }\Eprint
  {http://arxiv.org/abs/2105.06470} {arXiv:2105.06470 [hep-ph]} \BibitemShut
  {NoStop}%
\bibitem [{\citenamefont {Fischer}\ \emph {et~al.}(2020)\citenamefont
  {Fischer}, \citenamefont {Hern\'andez-Cabezudo},\ and\ \citenamefont
  {Schwetz}}]{Fischer:2019fbw}%
  \BibitemOpen
  \bibfield  {author} {\bibinfo {author} {\bibfnamefont {O.}~\bibnamefont
  {Fischer}}, \bibinfo {author} {\bibfnamefont {A.}~\bibnamefont
  {Hern\'andez-Cabezudo}}, \ and\ \bibinfo {author} {\bibfnamefont
  {T.}~\bibnamefont {Schwetz}},\ }\href {\doibase 10.1103/PhysRevD.101.075045}
  {\bibfield  {journal} {\bibinfo  {journal} {Phys. Rev. D}\ }\textbf {\bibinfo
  {volume} {101}},\ \bibinfo {pages} {075045} (\bibinfo {year} {2020})},\
  \Eprint {http://arxiv.org/abs/1909.09561} {arXiv:1909.09561 [hep-ph]}
  \BibitemShut {NoStop}%
\bibitem [{\citenamefont {Jordan}\ \emph {et~al.}(2019)\citenamefont {Jordan},
  \citenamefont {Kahn}, \citenamefont {Krnjaic}, \citenamefont {Moschella},\
  and\ \citenamefont {Spitz}}]{Jordan:2018qiy}%
  \BibitemOpen
  \bibfield  {author} {\bibinfo {author} {\bibfnamefont {J.~R.}\ \bibnamefont
  {Jordan}}, \bibinfo {author} {\bibfnamefont {Y.}~\bibnamefont {Kahn}},
  \bibinfo {author} {\bibfnamefont {G.}~\bibnamefont {Krnjaic}}, \bibinfo
  {author} {\bibfnamefont {M.}~\bibnamefont {Moschella}}, \ and\ \bibinfo
  {author} {\bibfnamefont {J.}~\bibnamefont {Spitz}},\ }\href {\doibase
  10.1103/PhysRevLett.122.081801} {\bibfield  {journal} {\bibinfo  {journal}
  {Phys. Rev. Lett.}\ }\textbf {\bibinfo {volume} {122}},\ \bibinfo {pages}
  {081801} (\bibinfo {year} {2019})},\ \Eprint
  {http://arxiv.org/abs/1810.07185} {arXiv:1810.07185 [hep-ph]} \BibitemShut
  {NoStop}%
\bibitem [{\citenamefont {Abratenko}\ \emph
  {et~al.}({\natexlab{e}})\citenamefont {Abratenko} \emph {et~al.}}]{gLEE_PRL}%
  \BibitemOpen
  \bibfield  {author} {\bibinfo {author} {\bibfnamefont {P.}~\bibnamefont
  {Abratenko}} \emph {et~al.} (\bibinfo {collaboration} {MicroBooNE}),\
  }\Eprint {http://arxiv.org/abs/2110.00409} {arXiv:2110.00409 [hep-ex]}
  \BibitemShut {NoStop}%
\bibitem [{\citenamefont {Acciarri}\ \emph
  {et~al.}(2017{\natexlab{e}})\citenamefont {Acciarri} \emph
  {et~al.}}]{Acciarri_2017}%
  \BibitemOpen
  \bibfield  {author} {\bibinfo {author} {\bibfnamefont {R.}~\bibnamefont
  {Acciarri}} \emph {et~al.} (\bibinfo {collaboration} {MicroBooNE}),\ }\href
  {\doibase 10.1088/1748-0221/12/02/P02017} {\bibfield  {journal} {\bibinfo
  {journal} {JINST}\ }\textbf {\bibinfo {volume} {12}},\ \bibinfo {pages}
  {P02017} (\bibinfo {year} {2017}{\natexlab{e}})},\ \Eprint
  {http://arxiv.org/abs/1612.05824} {arXiv:1612.05824 [physics.ins-det]}
  \BibitemShut {NoStop}%
\bibitem [{\citenamefont {Adams}\ \emph
  {et~al.}(2019{\natexlab{d}})\citenamefont {Adams} \emph {et~al.}}]{uB-CRT}%
  \BibitemOpen
  \bibfield  {author} {\bibinfo {author} {\bibfnamefont {C.}~\bibnamefont
  {Adams}} \emph {et~al.} (\bibinfo {collaboration} {MicroBooNE}),\ }\href
  {\doibase 10.1088/1748-0221/14/04/P04004} {\bibfield  {journal} {\bibinfo
  {journal} {JINST}\ }\textbf {\bibinfo {volume} {14}},\ \bibinfo {pages}
  {P04004} (\bibinfo {year} {2019}{\natexlab{d}})},\ \Eprint
  {http://arxiv.org/abs/1901.02862} {arXiv:1901.02862 [physics.ins-det]}
  \BibitemShut {NoStop}%
\bibitem [{\citenamefont {Abratenko}\ \emph
  {et~al.}({\natexlab{f}})\citenamefont {Abratenko} \emph {et~al.}}]{DL_PRD}%
  \BibitemOpen
  \bibfield  {author} {\bibinfo {author} {\bibfnamefont {P.}~\bibnamefont
  {Abratenko}} \emph {et~al.} (\bibinfo {collaboration} {MicroBooNE}),\
  }\Eprint {http://arxiv.org/abs/2110.14080} {arXiv:2110.14080 [hep-ex]}
  \BibitemShut {NoStop}%
\bibitem [{\citenamefont {Abratenko}\ \emph
  {et~al.}({\natexlab{g}})\citenamefont {Abratenko} \emph
  {et~al.}}]{PeLEE_PRD}%
  \BibitemOpen
  \bibfield  {author} {\bibinfo {author} {\bibfnamefont {P.}~\bibnamefont
  {Abratenko}} \emph {et~al.} (\bibinfo {collaboration} {MicroBooNE}),\
  }\Eprint {http://arxiv.org/abs/2110.14065} {arXiv:2110.14065 [hep-ex]}
  \BibitemShut {NoStop}%
\bibitem [{\citenamefont {Abratenko}\ \emph
  {et~al.}({\natexlab{h}})\citenamefont {Abratenko} \emph {et~al.}}]{WC_PRD}%
  \BibitemOpen
  \bibfield  {author} {\bibinfo {author} {\bibfnamefont {P.}~\bibnamefont
  {Abratenko}} \emph {et~al.} (\bibinfo {collaboration} {MicroBooNE}),\
  }\Eprint {http://arxiv.org/abs/2110.13978} {arXiv:2110.13978 [hep-ex]}
  \BibitemShut {NoStop}%
\bibitem [{\citenamefont {Agostinelli}\ \emph {et~al.}(2003)\citenamefont
  {Agostinelli} \emph {et~al.}}]{GEANT4:2002zbu}%
  \BibitemOpen
  \bibfield  {author} {\bibinfo {author} {\bibfnamefont {S.}~\bibnamefont
  {Agostinelli}} \emph {et~al.} (\bibinfo {collaboration} {GEANT4}),\ }\href
  {\doibase 10.1016/S0168-9002(03)01368-8} {\bibfield  {journal} {\bibinfo
  {journal} {Nucl. Instrum. Meth. A}\ }\textbf {\bibinfo {volume} {506}},\
  \bibinfo {pages} {250} (\bibinfo {year} {2003})}\BibitemShut {NoStop}%
\bibitem [{\citenamefont {Aguilar-Arevalo}\ \emph {et~al.}(2009)\citenamefont
  {Aguilar-Arevalo} \emph {et~al.}}]{Aguilar_Arevalo_2009}%
  \BibitemOpen
  \bibfield  {author} {\bibinfo {author} {\bibfnamefont {A.~A.}\ \bibnamefont
  {Aguilar-Arevalo}} \emph {et~al.} (\bibinfo {collaboration} {MiniBooNE}),\
  }\href {\doibase 10.1103/PhysRevD.79.072002} {\bibfield  {journal} {\bibinfo
  {journal} {Phys. Rev. D}\ }\textbf {\bibinfo {volume} {79}},\ \bibinfo
  {pages} {072002} (\bibinfo {year} {2009})},\ \Eprint
  {http://arxiv.org/abs/0806.1449} {arXiv:0806.1449 [hep-ex]} \BibitemShut
  {NoStop}%
\bibitem [{\citenamefont {Andreopoulos}\ \emph {et~al.}(2010)\citenamefont
  {Andreopoulos} \emph {et~al.}}]{Andreopoulos_2010}%
  \BibitemOpen
  \bibfield  {author} {\bibinfo {author} {\bibfnamefont {C.}~\bibnamefont
  {Andreopoulos}} \emph {et~al.},\ }\href {\doibase 10.1016/j.nima.2009.12.009}
  {\bibfield  {journal} {\bibinfo  {journal} {Nucl. Instrum. Meth. A}\ }\textbf
  {\bibinfo {volume} {614}},\ \bibinfo {pages} {87} (\bibinfo {year} {2010})},\
  \Eprint {http://arxiv.org/abs/0905.2517} {arXiv:0905.2517 [hep-ph]}
  \BibitemShut {NoStop}%
\bibitem [{\citenamefont {Abratenko}\ \emph
  {et~al.}({\natexlab{i}})\citenamefont {Abratenko} \emph
  {et~al.}}]{genie-tune-paper}%
  \BibitemOpen
  \bibfield  {author} {\bibinfo {author} {\bibfnamefont {P.}~\bibnamefont
  {Abratenko}} \emph {et~al.} (\bibinfo {collaboration} {MicroBooNE}),\
  }\Eprint {http://arxiv.org/abs/2110.14028} {arXiv:2110.14028 [hep-ex]}
  \BibitemShut {NoStop}%
\bibitem [{\citenamefont {Church}\ \emph {et~al.}()\citenamefont {Church} \emph
  {et~al.}}]{larsoft}%
  \BibitemOpen
  \bibfield  {author} {\bibinfo {author} {\bibfnamefont {E.}~\bibnamefont
  {Church}} \emph {et~al.} (\bibinfo {collaboration} {LArSoft}),\ }\bibinfo
  {note} {\url{https://larsoft.org}},\ \Eprint {http://arxiv.org/abs/1311.6774}
  {arXiv:1311.6774 [physics.ins-det]} \BibitemShut {NoStop}%
\bibitem [{\citenamefont {Abratenko}\ \emph
  {et~al.}(2021{\natexlab{j}})\citenamefont {Abratenko} \emph
  {et~al.}}]{wire-mod-paper}%
  \BibitemOpen
  \bibfield  {author} {\bibinfo {author} {\bibfnamefont {P.}~\bibnamefont
  {Abratenko}} \emph {et~al.} (\bibinfo {collaboration} {MicroBooNE})\
  }(\bibinfo {year} {2021})\ \Eprint {http://arxiv.org/abs/2111.03556}
  {arXiv:2111.03556 [hep-ex]} \BibitemShut {NoStop}%
\bibitem [{\citenamefont {Ji}\ \emph {et~al.}(2020)\citenamefont {Ji},
  \citenamefont {Gu}, \citenamefont {Qian}, \citenamefont {Wei},\ and\
  \citenamefont {Zhang}}]{Ji:2019yca}%
  \BibitemOpen
  \bibfield  {author} {\bibinfo {author} {\bibfnamefont {X.}~\bibnamefont
  {Ji}}, \bibinfo {author} {\bibfnamefont {W.}~\bibnamefont {Gu}}, \bibinfo
  {author} {\bibfnamefont {X.}~\bibnamefont {Qian}}, \bibinfo {author}
  {\bibfnamefont {H.}~\bibnamefont {Wei}}, \ and\ \bibinfo {author}
  {\bibfnamefont {C.}~\bibnamefont {Zhang}},\ }\href {\doibase
  10.1016/j.nima.2020.163677} {\bibfield  {journal} {\bibinfo  {journal} {Nucl.
  Instrum. Meth. A}\ }\textbf {\bibinfo {volume} {961}},\ \bibinfo {pages}
  {163677} (\bibinfo {year} {2020})},\ \Eprint
  {http://arxiv.org/abs/1903.07185} {arXiv:1903.07185 [physics.data-an]}
  \BibitemShut {NoStop}%
\bibitem [{\citenamefont {Abratenko}\ \emph
  {et~al.}(2021{\natexlab{k}})\citenamefont {Abratenko} \emph
  {et~al.}}]{MicroBooNE:2020sar}%
  \BibitemOpen
  \bibfield  {author} {\bibinfo {author} {\bibfnamefont {P.}~\bibnamefont
  {Abratenko}} \emph {et~al.} (\bibinfo {collaboration} {MicroBooNE}),\ }\href
  {\doibase 10.1088/1748-0221/16/02/P02017} {\bibfield  {journal} {\bibinfo
  {journal} {JINST}\ }\textbf {\bibinfo {volume} {16}},\ \bibinfo {pages}
  {P02017} (\bibinfo {year} {2021}{\natexlab{k}})},\ \Eprint
  {http://arxiv.org/abs/2002.09375} {arXiv:2002.09375 [physics.ins-det]}
  \BibitemShut {NoStop}%
\bibitem [{\citenamefont {Abratenko}\ \emph
  {et~al.}({\natexlab{j}})\citenamefont {Abratenko} \emph
  {et~al.}}]{DL_shower}%
  \BibitemOpen
  \bibfield  {author} {\bibinfo {author} {\bibfnamefont {P.}~\bibnamefont
  {Abratenko}} \emph {et~al.} (\bibinfo {collaboration} {MicroBooNE}),\
  }\Eprint {http://arxiv.org/abs/2110.11874} {arXiv:2110.11874 [hep-ex]}
  \BibitemShut {NoStop}%
\bibitem [{\citenamefont {Abratenko}\ \emph
  {et~al.}(2021{\natexlab{l}})\citenamefont {Abratenko} \emph
  {et~al.}}]{MicroBooNE:2020yze}%
  \BibitemOpen
  \bibfield  {author} {\bibinfo {author} {\bibfnamefont {P.}~\bibnamefont
  {Abratenko}} \emph {et~al.} (\bibinfo {collaboration} {MicroBooNE}),\ }\href
  {\doibase 10.1103/PhysRevD.103.052012} {\bibfield  {journal} {\bibinfo
  {journal} {Phys. Rev. D}\ }\textbf {\bibinfo {volume} {103}},\ \bibinfo
  {pages} {052012} (\bibinfo {year} {2021}{\natexlab{l}})},\ \Eprint
  {http://arxiv.org/abs/2012.08513} {arXiv:2012.08513 [physics.ins-det]}
  \BibitemShut {NoStop}%
\bibitem [{\citenamefont {Abratenko}\ \emph
  {et~al.}(2021{\natexlab{m}})\citenamefont {Abratenko} \emph
  {et~al.}}]{MicroBooNE:2020hho}%
  \BibitemOpen
  \bibfield  {author} {\bibinfo {author} {\bibfnamefont {P.}~\bibnamefont
  {Abratenko}} \emph {et~al.} (\bibinfo {collaboration} {MicroBooNE}),\ }\href
  {\doibase 10.1103/PhysRevD.103.092003} {\bibfield  {journal} {\bibinfo
  {journal} {Phys. Rev. D}\ }\textbf {\bibinfo {volume} {103}},\ \bibinfo
  {pages} {092003} (\bibinfo {year} {2021}{\natexlab{m}})},\ \Eprint
  {http://arxiv.org/abs/2010.08653} {arXiv:2010.08653 [hep-ex]} \BibitemShut
  {NoStop}%
\bibitem [{\citenamefont {Abratenko}\ \emph
  {et~al.}({\natexlab{k}})\citenamefont {Abratenko} \emph {et~al.}}]{ub_pid}%
  \BibitemOpen
  \bibfield  {author} {\bibinfo {author} {\bibfnamefont {P.}~\bibnamefont
  {Abratenko}} \emph {et~al.} (\bibinfo {collaboration} {MicroBooNE}),\
  }\Eprint {http://arxiv.org/abs/2109.02460} {arXiv:2109.02460
  [physics.ins-det]} \BibitemShut {NoStop}%
\bibitem [{\citenamefont {Adamson}\ \emph {et~al.}(2016)\citenamefont {Adamson}
  \emph {et~al.}}]{Adamson:2015dkw}%
  \BibitemOpen
  \bibfield  {author} {\bibinfo {author} {\bibfnamefont {P.}~\bibnamefont
  {Adamson}} \emph {et~al.},\ }\href {\doibase 10.1016/j.nima.2015.08.063}
  {\bibfield  {journal} {\bibinfo  {journal} {Nucl. Instrum. Meth. A}\ }\textbf
  {\bibinfo {volume} {806}},\ \bibinfo {pages} {279} (\bibinfo {year}
  {2016})},\ \Eprint {http://arxiv.org/abs/1507.06690} {arXiv:1507.06690
  [physics.acc-ph]} \BibitemShut {NoStop}%
\bibitem [{\citenamefont {Qian}\ \emph {et~al.}(2018)\citenamefont {Qian},
  \citenamefont {Zhang}, \citenamefont {Viren},\ and\ \citenamefont
  {Diwan}}]{Qian:2018qbv}%
  \BibitemOpen
  \bibfield  {author} {\bibinfo {author} {\bibfnamefont {X.}~\bibnamefont
  {Qian}}, \bibinfo {author} {\bibfnamefont {C.}~\bibnamefont {Zhang}},
  \bibinfo {author} {\bibfnamefont {B.}~\bibnamefont {Viren}}, \ and\ \bibinfo
  {author} {\bibfnamefont {M.}~\bibnamefont {Diwan}},\ }\href {\doibase
  10.1088/1748-0221/13/05/P05032} {\bibfield  {journal} {\bibinfo  {journal}
  {JINST}\ }\textbf {\bibinfo {volume} {13}},\ \bibinfo {pages} {P05032}
  (\bibinfo {year} {2018})},\ \Eprint {http://arxiv.org/abs/1803.04850}
  {arXiv:1803.04850 [physics.ins-det]} \BibitemShut {NoStop}%
\bibitem [{\citenamefont {Abratenko}\ \emph
  {et~al.}(2021{\natexlab{n}})\citenamefont {Abratenko} \emph
  {et~al.}}]{MicroBooNE:2020vry}%
  \BibitemOpen
  \bibfield  {author} {\bibinfo {author} {\bibfnamefont {P.}~\bibnamefont
  {Abratenko}} \emph {et~al.} (\bibinfo {collaboration} {MicroBooNE}),\ }\href
  {\doibase 10.1088/1748-0221/16/06/P06043} {\bibfield  {journal} {\bibinfo
  {journal} {JINST}\ }\textbf {\bibinfo {volume} {16}},\ \bibinfo {pages}
  {P06043} (\bibinfo {year} {2021}{\natexlab{n}})},\ \Eprint
  {http://arxiv.org/abs/2011.01375} {arXiv:2011.01375 [physics.ins-det]}
  \BibitemShut {NoStop}%
\bibitem [{\citenamefont {Abratenko}\ \emph
  {et~al.}({\natexlab{l}})\citenamefont {Abratenko} \emph
  {et~al.}}]{wc_pattern_recognition}%
  \BibitemOpen
  \bibfield  {author} {\bibinfo {author} {\bibfnamefont {P.}~\bibnamefont
  {Abratenko}} \emph {et~al.} (\bibinfo {collaboration} {MicroBooNE}),\
  }\Eprint {http://arxiv.org/abs/2110.13961} {arXiv:2110.13961 [hep-ex]}
  \BibitemShut {NoStop}%
\bibitem [{\citenamefont {Abratenko}\ \emph
  {et~al.}({\natexlab{m}})\citenamefont {Abratenko} \emph
  {et~al.}}]{MB-LEE-model}%
  \BibitemOpen
  \bibfield  {author} {\bibinfo {author} {\bibfnamefont {P.}~\bibnamefont
  {Abratenko}} \emph {et~al.} (\bibinfo {collaboration} {MicroBooNE}),\
  }\bibinfo {note}
  {\url{https://microboone.fnal.gov/wp-content/uploads/MICROBOONE-NOTE-1043-PUB.pdf}}\BibitemShut
  {NoStop}%
\bibitem [{\citenamefont {D'Agostini}(1995)}]{DAgostini:1994fjx}%
  \BibitemOpen
  \bibfield  {author} {\bibinfo {author} {\bibfnamefont {G.}~\bibnamefont
  {D'Agostini}},\ }\href {\doibase 10.1016/0168-9002(95)00274-X} {\bibfield
  {journal} {\bibinfo  {journal} {Nucl. Instrum. Meth. A}\ }\textbf {\bibinfo
  {volume} {362}},\ \bibinfo {pages} {487} (\bibinfo {year}
  {1995})}\BibitemShut {NoStop}%
\bibitem [{\citenamefont {Aguilar-Arevalo}\ \emph {et~al.}(2018)\citenamefont
  {Aguilar-Arevalo} \emph {et~al.}}]{MiniBooNE:2018esg}%
  \BibitemOpen
  \bibfield  {author} {\bibinfo {author} {\bibfnamefont {A.~A.}\ \bibnamefont
  {Aguilar-Arevalo}} \emph {et~al.} (\bibinfo {collaboration} {MiniBooNE}),\
  }\href {\doibase 10.1103/PhysRevLett.121.221801} {\bibfield  {journal}
  {\bibinfo  {journal} {Phys. Rev. Lett.}\ }\textbf {\bibinfo {volume} {121}},\
  \bibinfo {pages} {221801} (\bibinfo {year} {2018})},\ \Eprint
  {http://arxiv.org/abs/1805.12028} {arXiv:1805.12028 [hep-ex]} \BibitemShut
  {NoStop}%
\bibitem [{\citenamefont {Feldman}\ and\ \citenamefont
  {Cousins}(1998)}]{Feldman:1997qc}%
  \BibitemOpen
  \bibfield  {author} {\bibinfo {author} {\bibfnamefont {G.~J.}\ \bibnamefont
  {Feldman}}\ and\ \bibinfo {author} {\bibfnamefont {R.~D.}\ \bibnamefont
  {Cousins}},\ }\href {\doibase 10.1103/PhysRevD.57.3873} {\bibfield  {journal}
  {\bibinfo  {journal} {Phys. Rev. D}\ }\textbf {\bibinfo {volume} {57}},\
  \bibinfo {pages} {3873} (\bibinfo {year} {1998})},\ \Eprint
  {http://arxiv.org/abs/physics/9711021} {arXiv:physics/9711021} \BibitemShut
  {NoStop}%
\bibitem [{\citenamefont {Antonello}\ \emph {et~al.}()\citenamefont {Antonello}
  \emph {et~al.}}]{MicroBooNE:2015bmn}%
  \BibitemOpen
  \bibfield  {author} {\bibinfo {author} {\bibfnamefont {M.}~\bibnamefont
  {Antonello}} \emph {et~al.} (\bibinfo {collaboration} {MicroBooNE, LAr1-ND,
  ICARUS-WA104}),\ }\Eprint {http://arxiv.org/abs/1503.01520} {arXiv:1503.01520
  [physics.ins-det]} \BibitemShut {NoStop}%
\end{thebibliography}%
\end{document}